\documentstyle[11pt,emulateapj]{article}

\tighten

\begin{document}

\def\beginrefer{\section*{References}%
\begin{quotation}\mbox{}\par}
\def\refer#1\par{{\setlength{\parindent}{-\leftmargin}\indent#1\par}}
\def\endrefer{\end{quotation}}

\submitted{The Astrophysical Journal, {\it accepted}}
\title{The Chandra X-ray Observatory Resolves the X-ray Morphology and 
Spectra of a Jet in PKS~0637-752}

\author{G. Chartas\altaffilmark{1}, D. M. Worrall\altaffilmark{3} \altaffilmark{4}, 
M. Birkinshaw\altaffilmark{3} \altaffilmark{4}, M. Cresitello-Dittmar\altaffilmark{4}, W. Cui\altaffilmark{2}, 
K. K. Ghosh\altaffilmark{5}, D. E. Harris\altaffilmark{4}, E. J. Hooper\altaffilmark{4}, D. L. Jauncey\altaffilmark{6},  
D. - W. Kim\altaffilmark{4}, J. Lovell\altaffilmark{6},
H. L. Marshall\altaffilmark{2}, S. Mathur\altaffilmark{7}, D. A. Schwartz\altaffilmark{4},
S. J. Tingay\altaffilmark{8}, S. N. Virani\altaffilmark{4}, 
and B. J. Wilkes\altaffilmark{4}}

\altaffiltext{1}{Astronomy and Astrophysics Department, Pennsylvania State University,
University Park, PA 16802., chartas@astro.psu.edu}
\altaffiltext{2}{MIT Center for Space Research, 70 Vassar Street, Cambridge, MA, 02139.}
\altaffiltext{3}{Department of Physics, University of Bristol, England, UK.}
\altaffiltext{4}{Harvard-Smithsonian Center For Astrophysics, Cambridge, MA 02138}
\altaffiltext{5}{Space Sciences Laboratory, NASA/Marshall Space Flight Center, Mail Code
ES84, Huntsville, AL 35812}
\altaffiltext{6}{Australia Telescope National facility, P.O. Box 76, 
Epping, NSW 2121, Australia.}
\altaffiltext{7}{The Ohio State University}
\altaffiltext{8}{Jet Propulsion Laboratory, California Institute of 
Technology, Mail Stop 238-332, 4800 Oak Grove Drive, Pasadena, CA 91109}

\begin{abstract}

The core-dominated radio-loud quasar PKS 0637-752 ($z = 0.654$) 
was the first celestial object observed with the Chandra X-ray Observatory,
offering the early surprise of the detection of a remarkable X-ray jet.
Several observations with a variety of detector
configurations contribute to a total exposure time with the Chandra
Advanced CCD Imaging Spectrometer (ACIS; Garmire et al. 2000, in preparation) of about 100~ks.
A spatial analysis of all the available X-ray data, 
making use of Chandra's spatial resolving power
of about 0.4 arcsec, reveals a jet that extends about 10 arcsec to the
west of the nucleus. At least four X-ray knots are resolved along the jet, 
which contains about 5\% of the overall X-ray luminosity of the
source. Previous observations of PKS~0637-752 in the radio band (Tingay et al. 1998)
had identified a kpc-scale radio jet extending to the West of the quasar.
The X-ray and radio jets are similar in shape, intensity
distribution, and angular structure out to about 9 arcsec, after which
the X-ray brightness decreases more rapidly and the radio jet turns
abruptly to the north. The X-ray luminosity of the total source is
$\log L_X \approx 45.8 \ \rm erg \, s^{-1} \ (2 - 10 \ keV)$, 
\footnote{We use $H_{0}$ = 50 km s$^{-1}$ Mpc$^{-1}$ 
and $q_{0}$ = 0 throughout}  
and appears not to have changed since it was observed with ASCA in
November 1996. We present the results of fitting a variety of emission
models to the observed spectral distribution,
comment on the non-existence of emission lines
recently reported in the ASCA observations of PKS~0637-752,
and briefly discuss plausible X-ray emission mechanisms.

\end{abstract}

\section{INTRODUCTION}
Until recently most of our knowledge regarding the spatial structure
and spectral shape of extragalactic jets has relied on observations
performed in the radio and optical bands. 
X-ray detections and upper limits have provided insights into the
physical conditions responsible for the observed radiation 
from knots and hotspots in extragalactic radio jets. 
However, in most cases the poor spectral and spatial resolution 
available has made the interpretation of the X-ray data difficult.
The spectral energy distributions (SED) seen from jets have been ascribed to combinations of
synchrotron radiation, synchrotron self-Compton (SSC) radiation,
and thermal bremsstrahlung from shock-heated gas near the jets.
In particular, the synchrotron and SSC models have been successfully
employed to explain the X-ray emission from hotspots and jets
in M87 (Biretta, Stern, $\&$ Harris 1991), Cygnus A (Harris, Carilli, \&
Perley 1994) and 3C~295 (Harris et al., 2000). In the cases of 3C~273 
(Harris \& Stern 1987) and  Pictor A (Meisenheimer et al. 1989),
none of the standard processes yield satisfactory results.

The Chandra X-ray Observatory (CXO), launched on 1999 July 23, provides
a significant improvement over previous missions in combined
spatial and spectral resolution 
(see Weisskopf and O'Dell 1997, van Speybroeck et al.~1997)
which we expect will result in a significant increase
in the number of detected and resolved X-ray jets.
PKS~0637-752 is the first X-ray jet to have been discovered by 
{\it Chandra}. 
The quasar was originally detected in X-rays with the Einstein Observatory
(Elvis \& Fabbiano 1984).
Since then a peculiar emission-line feature at $\sim 0.97$~keV
was claimed in an ASCA SIS observation of the source (Yaqoob et al. 1998).
One of two Ginga observations of PKS 0637-752 has been
reported as showing a marginal detection of an Fe $\rm K\alpha$
line with an equivalent width of $103 \pm 85 \ \rm eV$ ($\pm 1\sigma$ errors; Lawson \& Turner 1997).

In $\S$ 2 we describe the spectral and spatial analysis of the core 
and jet, and our observations of the radio jet. The properties of sources in the vicinity of 
PKS~0637-752 are also briefly presented. Section 3 contains a brief description
of our attempt to apply standard jet models to the observed SED
of PKS~0637-752. We provide a thorough investigation of the underlying
emission processes in a companion paper (Schwartz et al. 2000, in preparation)
where more complex models are considered.

\section{DATA REDUCTION AND ANALYSIS}

Twenty-five observations of PKS~0637-752 were made
between 1999 August 14 and August 24 with the {\it Chandra} ACIS-S 
during the orbital activation and checkout phase of the mission. 
Table 1 summarizes several ACIS configuration parameters,
50$\%$ encircled energy radii, exposure times and estimated net count rates
over the entire ACIS band  
corresponding to each individual observation.

The primary purpose of these observations
was to focus the ACIS-S detector (in $\rm SIM_x$) and to
determine the position of the Chandra mirror
optical axis by pointing to PKS~0637-752 at different
$\rm (Src_y, Src_z)$ offsets. (Definitions for Src$_{y}$,
Src$_{z}$ and SIM$_{x}$ are presented in the notes of Table 1.)

The spectral and spatial analysis of the data is complicated
by the numerous configurations used during the observation of PKS~0637-752.
The point spread function (PSF) corresponding to each observation 
differs significantly  because of the nature of the 
calibration being performed. In particular, the 50$\%$ encircled 
energy radii varied between 0.4$''$ at best focus and about 1$''$
at 3$'$ off-axis pointing or poor focus (see Figure 1 and Table 1). 
The jet ($\S$ 2.1) was spatially resolved from the core 
and made little impact on the determination of the
best focus and best optical axis location.

A second important property of the data that we have accounted 
for in our analysis is pile-up  (see $\S$ 6.17 of the {\it Chandra Observatory Guide}).
Whenever the separation of two or more X-ray photons incident 
on a CCD is less than a few CCD pixels, and their arrival time  
lies within the same CCD frame readout, 
the CCD electronics may regard them 
%(depending on the selected event and split-event thresholds and detection cell size) 
as a single event with an amplitude given by the sum of the electron charge in
the 3x3 neighborhood of the pixel with the maximum 
detected charge.
A detected CCD event is characterized by the 
total charge within the 3x3 island and the distribution
of the charge (often referred to as the grade of the event) within that island. 
Pile-up may alter the grades and charges of events, thus 
affecting both their spatial and spectral distribution.
A manifestation of pile-up in observed spectra may be a 
reduction of detected events, spectral hardening 
of the continuum component and the apparent distortion
of the PSF of point-like objects.

The presence of pile-up is apparent in the observed spectra of the
core of PKS~0637-752.
In particular, we observed a significant change in 
count rate of the core component of PKS~0637-752 as a function 
of focus position ($\rm SIM_x$) along the optical axis.
As shown in figure 1, the apparent count rate is lowest at the best focus
position and increases as the scientific instrument module (SIM) is moved away from this location.
A parabolic function was used to fit the count rates and encircled energy radii 
as a function of focus position.
We estimate a $\rm SIM_x$ focus location of $\sim$ -0.7~mm corresponding to the minimum
of both count rate and encircled energy radii. 

Count rates for a similar test performed with a shorter CCD
frame readout time showed a similar behavior, but 
with detected count rates systematically larger than those
for the standard 3.24~s full-frame readout time.
The observed count-rate variation is due to the pile-up effect
and not intrinsic to the quasar. As the telescope approaches
best focus most photons fall onto a single CCD 
pixel, and the enhanced pile-up leads to a decrease in detected count rate.

Correcting a CCD observation of an X-ray source for 
pile-up is quite complicated. The dependence of pile-up on 
several effects such as the incident X-ray flux, the 
spectral and spatial distribution of the events, the 
grade selection scheme used, and the detection cell size adopted
%and the event and split-event thresholds 
make the problem of restoring to the unpiled up spectrum non-trivial.    

Fortunately the jet of PKS~0637-752 is not 
affected by pile-up since the jet is a relatively low
count rate extended source.

\subsection{X-ray Morphology} To obtain a high signal-to-noise image of PKS~0637-752, 
we made use of all observations with observed half power radius (HPR) less than 1.2$''$. 
For each observation, a PSF appropriate for 
the focus and aim-point position was created employing the simulation tool MARX v2.2
(Wise et al. 1997).
The input spectrum assumed in the PSF simulations was that
derived from the best-fit Chandra spectrum of the outer-jet of PKS~0637-752 (see section 2.3). 
Specifically, we used an absorbed power law with a 
column density of N$_{H}$ = 11.8 $\times$ 10$^{20}$ cm$^{-2}$ 
and a photon index of 1.83. We examined the sensitivity 
of the resulting deconvolved image to the spectral slope
used in the PSF simulations and found no particular spectral 
dependence for photon indices ranging between 1.7 and 2.3.

The simulated PSF's were binned to a sub-pixel scale of 0.125$''$. 
%The aspect reconstruction process at the Chandra X-ray Center 
%(CXC) produces photon event positions in sky coordinates 
%which are not pixelized. 
To avoid aliasing effects, the Chandra X-ray Center (CXC) processing 
incorporates a randomization of each position by 
$\pm$ 0.246 arcsec (1 ACIS pixel = 0.492$''$). 
Residual errors in the aspect solution are expected to add 
a ``blurring'' to detected photon positions of $\sim$ 0.3$''$ RMS in diameter.
To simulate aspect errors and position randomization we convolved each generated PSF with a 
Gaussian with $\sigma$ = 0.25$''$. The X-ray photon event 
positions for each observation were also binned to 0.125$''$, 
and the resulting X-ray image is shown in Fig. 2(Top panel).
A maximum-likelihood  deconvolution technique, using the appropriate simulated PSF,
was applied to each individual observation. The resulting deconvolved
images were aligned on the core centroid and combined to produce the total deconvolved
image of PKS~0637-752 shown in Fig. 2(Middle panel). 
The effective resolution of the deconvolved X-ray image 
was estimated by deconvolving the simulated PSF's of
each observation, stacking the simulated PSF's
and determining the FWHM of the deconvolved stacked PSF.
Our analysis yields an effective resolution for the stacked deconvolved
image of $\sim$ 0.37$''$. The PSF's simulated for this analysis
are not appropriate for the piled-up region of the core. 
This is why any deconvolved structure within $\sim$ 2$''$ from the core should not be
considered real.  The jet region beyond $\sim$ 2$''$
does not suffer from pile-up effects and the 
simulated PSF's are appropriate for deconvolving the jet region.
Several interesting structures have 
become more apparent in Fig. 2(Middle panel). A well collimated X-ray jet is
seen to originate in the core and to extend approximately 10$''$ to the west,
and within the jet at least four knots are clearly resolved.

\subsection{Radio Morphology}
The radio jet was imaged at 4.8 and 8.6~GHz on 1999 August 19 and on 1999 September 21
using the Australia Telescope Compact Array which has a similar arcsecond
resolution at 8.6~GHz as Chandra, and so provides a powerful structural
comparison. The resulting 8.6~GHz radio image (Schwartz et al. 2000, in preparation;
Lovell et al. 2000) is shown in Figure 2(Lower panel), where
the coincidence with the X-ray knots is apparent.
The three furthest X-ray knots, WK7.8, WK8.9 and WK9.7,
appear to be embedded in extended and diffuse X-ray emission. 
The X-ray jet appears to bend SW after its encounter with the first 
X-ray knot, WK5.7, and then bend NW after the encounter with the fourth
X-ray knot WK9.7. The radio image of the PKS~0637-752 jet
(Figure 2, lower panel) indicates that the radio jet also
bends in the NW direction after its encounter with WK9.7.
The X-ray emission after knot WK9.7 drops substantially to 
become undetectable after several arcsec. 
A comparison between the X-ray and radio intensity profile along the jet
(figure 3) shows close alignment
of the X-ray and radio jets out to 10 arcseconds and
relatively well matched X-ray and radio knots at 
5.7, 7.8, 8.9 and 9.7 $''$ from the core. The radio profile in 
figure 3 was produced from the 8.6~GHz image (Schwartz et al. 2000, in preparation; Lovell et al. 2000).
The position angles of the radio and X-ray jets appear 
to be similar, as shown in figure 4 where we have
used the deconvolved X-ray image and the same 8.6~GHz radio map as before to
plot the position angle of the ridges of peak radio and X-ray
brightness along the jet with respect to the core of PKS~0637-752.

\subsection{Spectral Analysis of Core Component}
To determine the continuum spectral shape of the core 
component of PKS~0637-752, we considered only the two 
least piled-up on-axis observations corresponding to obsid's 472 and 476.
The expected percent loss of counts due to pile-up,
based on our simulations, is $\sim$ 12${\%}$. 
X-ray events were extracted within a circle centered on the core of 
PKS~0637-752 with a radius of 5$''$,
and events with standard ASCA grades 0,2,3,4,6 only were selected. The background was
determined by extracting events within annuli centered 
on the core with inner and outer radii of 15$''$ and 20$''$ respectively.
Spectra were binned to have a minimum of 20 counts per bin such that $\chi^{2}$
statistics can be used without low count corrections.

Spectral fits were performed 
using the standard software tool XSPEC (Arnaud 1996), and we compared these results to 
those provided by the simulator-based spectral 
fitting tool LYNX (Chartas et al. 2000, in preparation). 
A brief description of the LYNX fitting tool is provided in appendix A.
The telescope pointings are intentionally dithered for the
observations of PKS~0637-752 in a Lissajous pattern with an amplitude 
of about 16 arcsec such that small-scale non-uniformities in the CCD quantum efficiency
are averaged out. The selected aim-point of the telescope 
for most of the PKS~0637-752 observations
was set within 10$''$ from the boundary between amplifier nodes 0 and 1 of the S3 CCD
of ACIS, resulting in the source being dithered across the two nodes.
To account for the difference in amplifier gains between nodes 0 and 1, events detected from the 
core component of PKS~0637-752 were separated into two spectra containing events
detected from a single node, either 0 or 1.
For fitting the spectra with XSPEC we used the appropriate response 
and ancillary files provided by the CXC. 
All errors on best-fit spectral parameters quoted in this paper are at the 90$\%$
confidence level unless mentioned otherwise.

Specifically we used the Chandra Interactive Analysis of Observations (CIAO) 
v1.1 tools {\it mkrmf} and {\it mkarf} to 
generate response and ancillary files. 
The focal plane temperature was $-100$ C for these observations.
%The FITS embedded functions used for building the response files 
%for nodes 0 and 1 of CCD S3 were selected to correspond to 
%the focal plane temperature of $-100$ C, at which these observations 
%were made, and to the detector location of the source.
The exposure times of each observation are listed in Table 1.
A simple power-law plus cold, neutral absorber model was used to 
fit the core spectrum of PKS~0637-752. 
The best-fit parameters of the spectral fits are shown in Table 2.

%The available response matrices and ancillary files at the time of writing 
%appear to be uncertain at energies below $\sim$ 0.5~keV. This is illustrated in 
%fit 2 which was performed in the energy range 0.2-6.0~keV. 

Fit 2 of Table 2 was performed in the energy range 0.2-6.0~keV.
The reduced $\chi$$^{2}_{\nu}$ of 1.1 
for 235 degrees of freedom is acceptable in a statistical sense. However, the best-fit
value of 3.5 $\pm$ 0.5 $\times$ 10$^{20}$ cm$^{-2}$ for the column density is 
significantly below the previously observed ASCA value of 9 $\pm$ 3 $\times$ 10$^{20}$ cm$^{-2}$ 
and the Galactic column density of 9.1 $\times$ 10$^{20}$ cm$^{-2}$ (Dickey \& Lockman 1990). 
In fit 3 of Table 2 we fixed the neutral column density to the Galactic
value and obtained a reduced $\chi$$^{2}_{\nu}$ of 1.7 for 236 degrees of freedom.

We searched for systematic errors in the spectral fits of the core by repeating
 fit 3 of Table 2 allowing for the
lower-bound of the fitted energy range to vary between 0.2 and 1.5keV.
We find that the best-fit XSPEC value for the spectral slope
varies between 2.00 $\pm$ 0.04 and 1.8 $\pm$ 0.05 for the lower-bounds of the fit
ranging between 0.2 and 0.8keV respectively,
and remains constant at about 1.8 for the lower-bounds of the fit
ranging between 0.8 and 1.5keV.
Plausible explanations for the variation of the spectral slope with the 
lower-bound of the fit include pile-up effects of the spectrum of the core
and/or uncertainties in the available response and ancillary files
at energies below $\sim$ 0.5keV.
We therefore restricted the XSPEC fits of the core component to energies above 1keV with the neutral column density 
parameter held fixed at the Galactic value of N$_{H}$ = 9 $\times$ 10$^{20}$ cm$^{-2}$.
The fit within the $1.-6$~keV range yields a photon index of 1.76 $\pm$ 0.1 (fit 4 in Table 2).

Best-fit parameters to the core component were also obtained utilizing the simulator based tool,
LYNX, (fit 1 in Table 2) and restricting the energy range of the fit between 0.2 and 6.0~keV. 
This fit yields a column density of 11 $\pm$ 2 $\times$ 10$^{20}$ cm$^{-2}$ 
that is consistent with the Galactic value, and a photon index of 1.77 $\pm$ 0.1 that is 
consistent with Ginga and ASCA results (Williams et al. 1992; Lawson \& Turner 1997 
and Yaqoob et al. 1998). All X-ray photon indices, $\Gamma$, and energy indices, $\alpha_{E}$, 
follow the convention of flux density $\propto$ $\nu^{-(\Gamma-1)}$, where 
flux density is in units of erg cm$^{-2}$ s$^{-1}$ Hz$^{-1}$and $\Gamma$ = $\alpha_{E}$ + 1.

The difference between the LYNX and XSPEC fit results for the spectrum of the core
of PKS~0637-752 can be partially attributed to the presence
of pile-up in the spectrum of the core which is not modeled in the XSPEC fits.

In figure 5 we present the spectrum of the core component
of PKS~0637-752 corresponding to obsid 476, together with the best-fit 
model (fit 1 from Table 2) and the ratio of observed spectrum to model.
The estimated 2-10~keV flux of 2.2  $\times$ 10$^{-12}$ erg s$^{-1}$ cm$^{-2}$
(from fit 1 of Table 2) indicates that the core flux has not varied,
within the quoted error bars, since the 1996 November ASCA observations,
especially when one considers that the ASCA observed flux 
corresponds to all emission within a 3$'$ radius of the core.
X-ray fluxes quoted in this paper are not corrected for Galactic absorption.

We chose to improve the signal to noise for the detection of
faint emission lines in the spectrum of PKS 0637-752 by stacking
all the available spectra. In the summed spectrum, we expect that
pileup will distort the continuum shape and cause emission lines
to appear with overtone ``ghosts'' at higher energies.
We initially produced a model that best fit the continuum 
component of the stacked spectrum. The region in the vicinity 
of the mirror Ir-edges (see Table 3) was excluded from the fit. 
In figure 6 we plot the stacked spectrum of the core
component of PKS~0637-752 with the best-fit continuum component.
We also show the ratio of the observed to best-fit continuum model.
The best-fit model yields a reduced ${\chi}^{2}_{\nu}$ of 
0.99 for 147 degrees of freedom (dof).
This ratio suggests that no significant emission-line 
features are present. In particular, we do not detect an emission line
in the vicinity of 1~keV and at an equivalent
width (EW) $\sim$ 60~eV as suggested by a recent 
ASCA observation of PKS~0637-752 (Yaqoob et al. 1998).
An upper limit of 15~eV on EW, at the 90$\%$ confidence level, can be placed 
on any line with energies 0.8 - 1.2~keV (observed frame) and width of 0-0.2~keV.

There are several spectral features in observed ACIS spectra 
that do not originate from astrophysical sources but are produced 
from various sites within the Chandra/ACIS instrument. 
In Table 3 we list all known instrumental spectral features
and their origin. The significant instrumental features 
are the Al-L, C-K, N-K, O-K, and Al-K UV/Optical blocking filter absorption edges,
the N-K, O-K, and Si-K CCD absorption edges,
and the Ir-M HRMA absorption edges. In the spectra
of the core of PKS~0637-752 (see Figures 5 and 6) 
the main instrumental spectral features
that can be seen are the Ir-M HRMA absorption edge at 2.085~keV
and the O-K$\alpha$ ACIS absorption edge at 0.536~keV.

\subsection{Spectral Analysis of Inner-Jet Component}
For the purposes of this spectral analysis we have extracted 
the spectrum of the jet in a region 4$''$ to 6.5$''$ away from the core,
and define this as the inner-jet region.
Pile-up effects are negligible for the inner-jet component due to 
the extended nature of the emission
combined with the relatively low count rate of $\sim$ 4 $\times$ 10$^{-3}$ counts s$^{-1}$.
The X-ray image of PKS~0637-752  shown in figure 2
indicates a degree of curvature in the jet that appears to 
follow the spatial morphology observed in the radio (Schwartz et al. 2000, in preparation; Lovell et al. 2000).
The X-ray spectrum of the inner-jet component was produced
by extracting events from all the observations listed in Table 1. The
spatial extraction filter chosen was a rectangle with the
left lower corner set at (x$_{c}$ + 4$''$,y$_{c}$-1.5$''$)
and the right upper corner set at (x$_{c}$ + 6.5$''$,y$_{c}$+1.5$''$),
where x$_{c}$, y$_{c}$ are the centroid locations of the core
for each observation. In figure 7 we show the observed spectrum
of the inner-jet component with the best-fit model
assuming a power-law emission process (fit 3 Table 4).

Because of the present uncertainties with response matrices
below 0.5~keV for the S3 CCD, the spectral fits utilizing XSPEC
were performed in the 0.6 to 4~keV range, and fits 
utilizing the tool LYNX (which appears to be more reliable at energies below 1~keV)
were performed in the 0.2 to 4~keV range.
Model 1 incorporates a simple absorbed power-law model
and yields a photon index of $\Gamma$ = 2.0 $\pm$ 0.2(LYNX, fit 1 in Table 4)
or $\Gamma$ = 2.27 $\pm$ 0.3(XSPEC, fit 3 in Table 4). 
Model 2 incorporates a Raymond-Smith thermal plasma model and 
a best-fit column density consistent with the Galactic value,
and yields a best-fit temperature of 2.7 $\pm$ 0.2~keV(LYNX, fit 2 in Table 4)
or 2.29$_{-0.5}^{+1.0}$keV(XSPEC, fit 4 in Table 4) .
Abundances were fixed at 0.3 of the cosmic value.

An F-test between fits 1 and 2 indicates that 
neither model is significantly preferred over the other. 
The 2-10~keV X-ray luminosity of the inner-jet region, assuming 
an absorbed power-law model (model 1 in Table 4), is 0.27 $\times$ 10$^{44}$ erg s$^{-1}$.

\subsection{Spectral Analysis of Outer-Jet Component}

The X-ray spectra for the outer-jet component were extracted from circles
centered on knot WK8.9 with radii of 2.5$''$. Only standard ASCA grades 0,2,3,4,6
were included, and the background was determined by extracting events within 
annuli centered on the core with inner and outer radii of 15$''$ and 
25$''$, respectively. 
Pile-up effects are negligible for the outer-jet component due to the extended nature of the emission
combined with the relatively low count rate of 0.025 counts s$^{-1}$.
We combined a subset of the observations of PKS~0637-752 listed in Table 1
to produce spectra of total exposure 32,931s for node 0 of S3 and 70,700s for node 1 of S3.
The XSPEC and LYNX  spectral fits were performed in the 0.6-6~keV range
and 0.2-6~keV range, respectively. 
Plausible emission mechanisms for the production of the observed X-rays
from the knots are synchrotron self Compton,
and thermal bremsstrahlung emission from a compressed shocked medium.
To distinguish between possible emission mechanisms we
fit the composite knot spectrum with 
absorbed power-law and thermal-plasma models.
More complex models were not pursued due to the relatively low counts 
in the composite spectrum (see Table 1) and the present uncertainties in the 
low-energy instrumental response. Our results are presented
in Table 5. Spectral fits with thermal and power-law models provide similar reduced
$\chi^{2}_{\nu}$. The 2-10~keV luminosity of the outer-jet region,
assuming an absorbed power-law model (Fit 1 in table 5), is 2.2 $\times$ 10$^{44}$ erg s$^{-1}$.
No significant emission lines are detected in the outer-jet spectrum.
The composite spectrum for the outer jet of PKS~0637-752
with best-fit model (Fit 3 in table 5) is shown in figure 7.

\subsection{Properties of Sources in the Near Vicinity of PKS~0637-752}

Several relatively X-ray bright sources (count rates above 
2 $\times$ 10$^{-3}$ cnts s$^{-1}$) were detected on CCD S3 in the vicinity of PKS~0637-752. 
In Table 6 we list their coordinates, count rates and distances 
from the core of PKS~0637-752. We searched the USNO
catalog and found optical counterparts within 1 arcsec in only  
five out of the 12 sources. No counterparts were found in the 
NED and SIMBAD catalogs. \\

\section{DISCUSSION}

In figure 8 we present the SED of the WK7.8 knot of PKS~0637-752.
The radio observations of PKS~0637-752 were 
performed at ATCA at 4.8 and 8.6~GHz. 
The 4.8 and 8.6~GHz beam width is $\sim$ 2 and $\sim$ 1 arcsec FWHM, 
respectively. The spectral indices and flux densities of the resolved components
of the core, jet, and knots are presented in Table 7.
Values for the optical flux density were obtained 
from the recent Hubble Space Telescope WFPC2 observations 
(Schwartz et al. 2000, in preparation and references therein).

The SED shows that a single-component power-law
synchrotron model cannot explain the combined radio, optical and 
X-ray flux densities, since the optical lies far below a power-law
interpolation between the radio and X-ray measurements.
We also tested whether SSC emission or inverse Compton scattering of
cosmic microwave background (CMB) photons could explain the observed
X-ray emission.  The model components in Figure 8 are for the case
of equipartition between the magnetic-field and electron energy densities,
and assume a sphere of radius 0.15 arcsec and a power-law electron number
spectrum of slope 2.4 between 100~MeV and 230~GeV, steepening by
unity at 30~GeV due to energy losses. A detailed description of the SSC and inverse Compton calculations
and the assumptions made for the model parameters are presented in a companion
paper (Schwartz et al. 2000, in preparation).
We estimate an equipartition field, B$_{eq}$, of about 2 $\times$ 10$^{-4}$ Gauss.
Based on this B$_{eq}$ value both SSC and IC on the CMB under-predict the X-ray 
flux by several orders of magnitude.
Thermal models were also considered.
Assuming a plasma temperature of 10~keV, an emission volume of
4 $\times$ 10$^{-3}$ arcsec$^{3}$ and a 2-10~keV luminosity
of 1 $\times$ 10$^{42}$ erg s$^{-1}$, we estimate a plasma
density of about 1 cm$^{-3}$.  
The derived rotation measures (RM)
for radio waves propagating through such a dense plasma are quite large,
inconsistent with the recent radio ATCA observations at 
4.8 and 8.6~GHz (Schwartz et al. 2000, in preparation; Lovell et al. 2000). 
These observations show a RM $\sim$ 80 rad m$^{-2}$
in the core, but no Faraday rotation in the jet,
with an upper limit of $\pm$ 10 rad m$^{-2}$.
A contrived geometry where the jet collides with a giant molecular cloud in a companion 
galaxy producing thermal X-rays but, from our line-of-sight, the cloud is 
located behind the jet, may explain the non-detection of Faraday rotation in the jet.

An examination of the radio and X-ray brightnesses of different
parts of the main emission region in the jet suggests that the 
X-ray brightness to radio brightness ratio is remarkably constant
out to the last knot (WK9.7), where the X-rays are relatively
fainter. The interpretation of the change of X-ray brightness between
WK8.9 and WK9.7 depends, however, on the emission process: if the
X-radiation has a synchrotron origin, then the emitting electrons
must be locally accelerated, and the change in X-ray brightness of
WK9.7 would be telling us about changes in particle acceleration at
different points in the jet. If the X-rays have an inverse Compton
origin, then it is possible that the brightness change is entirely
due to aging of an electron population accelerated in WK8.9, but
then other difficulties in understanding the energetics of the
source must be faced (see Schwartz 2000 for further discussion).

\section{CONCLUSIONS}
Chandra's unique resolving-power capabilities opens a new era in X-ray astronomy.
We anticipate many more radio jets and knots will be resolved 
in future observations with Chandra. The simultaneous
spatial and spectral information provided
with the Chandra/ACIS combination allows for accurate 
estimates of the size and spectral densities of the knots 
in extragalactic radio jets,
which leads to tighter constraints on models 
that attempt to explain the X-ray emission. 
The standard emission processes usually invoked to
explain X-ray emission from jets cannot explain the 
X-ray observations of PKS~0637-752. In particular, simple synchrotron models  
and equipartition SSC models under-predict the X-ray flux of the knots by many orders of magnitude. 
Thermal models predict shocked plasma
densities and rotation measures that are too large.
Particular contrived geometries of the jet and interacting molecular cloud
may, however, explain the non-detection of Faraday rotation in the radio observations.
More complex models that invoke inhomogeneities and/or non equipartition and/or an extra photon 
source to explain the X-ray emission as inverse Compton are presented in the paper 
Schwartz et al. (2000), in preparation.\\

A careful spatial analysis combining most of the
available observations of PKS~0637-752
has resolved the jet and at least four knots along the jet.
The X-ray knots are 5.7, 7.8, 8.9 and 9.7 $''$ from the 
core of PKS~0637-752, in good agreement
with the locations seen in the radio image.
The X-ray knots are not individually resolved 
and the upper limit on their diameter is $\sim$ 0.4$''$.
The radio knots from the ATCA image are unresolved with an upper limit on
their diameter also of $\sim$0.4$''$. However, our VLBI observations (Schwartz et
al. 2000, in preparation; Lovell et al. 2000) have been reanalyzed to search for compact
components close to the radio knots, and we find that the knots are indeed
resolved at 0.05$''$ resolution at 5 GHz, with less than 5 mJy remaining at
this resolution. This suggests that they are low surface brightness
``hot-spots''.

The spectral analysis of the core, jet and knot components
has been quite complex due to the different non-optimal configurations 
used for each observation and the present uncertainty 
in several of the Chandra/ACIS calibration data sets.
Having quoted the above caveat we summarize
the spectral analysis of the core and jet as follows:\\ 

(1) The core flux and spectral shape are consistent with those measured
with ASCA. However, we do not detect the emission line near 1~keV claimed
in a recent ASCA observation of PKS~0637-752.
The HPR for ASCA is about 3 arcmin, so one possible explanation 
for this discrepancy is that the emission line claimed to be
detected with ASCA may originate from a nearby unresolved source. 
We have detected three relatively weak X-ray sources within 3 arcmin of PKS~0637-752. However, the 
spectra of these sources do not show any prominent 
1~keV emission lines which could explain 
the ASCA results. \\

(2) The X-ray spectrum of the inner-jet component appears 
to be slightly steeper than that of the outer-jet region (see Tables 4 and 5).
A difference in spectral slopes may be explained as follows: 
A population of synchrotron-emitting relativistic electrons 
in the inner-jet region are undergoing radiation losses and producing the observed
steeper spectra (X-ray and radio spectral indices $\alpha_{E}$ $\sim$ -1.0).
As these electrons enter the outer-jet region (6.5$''$ - 11.5$''$ away from core) 
they are re-accelerated in a shock or some other structure to
produce a flatter spectrum (X-ray and radio spectral indices $\alpha_{E}$ $\sim$ -0.8). 
As they age past 11.5$''$ the radio spectrum steepens again (radio spectral index $\alpha_{E}$ $\sim$ -1.0).
Also, in the region 11.5$''$ away from the core where the highest-energy 
electrons may plausibly have lost all their energy, the X-rays turn off as expected. \\

(3) Spectral fits to the outer-jet spectrum assuming thermal and power-law 
models yield similar $\chi^{2}$'s. The outer-jet region is relatively bright in X-rays
with a 2-10~keV luminosity of 2.2 $\times$ 10$^{44}$ erg s$^{-1}$ (fit 1 in Table 5).
The best-fit value of the X-ray luminosity of the outer-jet is not sensitive to the 
assumed emission process. The good agreement in X-ray and radio spectral slopes in the inner and
outer jet regions strongly suggests that there is substantial electron
acceleration in the knot complexes WK7.8 and WK8.9.   \\

We would like to thank Eric Feigelson for helpful comments, Martin Hardcastle 
for software used to generate figure 8 and Kenneth Lanzetta for providing 
the optical flux density of knot WK7.8 used in figure 8.
This work was supported by NASA grant NAS 8-38252. \\

\newpage

\small
\begin{centering}
APPENDIX A \\

{\it LYNX; A Simulated Based}

{\it Spectral Fitting Tool} 

\vspace{0.15in}

\end{centering}

\normalsize

Astrophysical X-ray spectra are commonly analyzed by creating parameterized
models for the incident spectra, folding these models through the
telescope and instrument responses, and then adjusting the
parameters by minimizing a metric such as $\chi^2$
formed between the observed and modeled spectra.
A tool that uses this approach and is widely used
to fit astrophysical X-ray spectra observed from a variety of
X-ray satellites is XSPEC. 
The telescope's effective-area dependence with off-axis angle
combined with the detector's quantum-efficiency dependence with energy are
usually incorporated into an auxiliary response file
while the response of a detector to mono-energetic photons of energy
E are incorporated into a spectral redistribution matrix.
The spectral redistribution matrix is usually created
from parameterizing the output of CCD simulations of
input mono-energetic spectra.
One approximation of forward fitting CCD spectra using
telescope and detector response matrices is that the
simulated CCD spectra are uniquely defined for a given model and
set of input model parameters. In reality, however,
several physical processes within CCD's are non-deterministic
such as fluorescent yields, absorption depths and photon escape
probabilities, and detected spectra
for identical incident spectra will in general be
slightly different. This effect becomes more noticeable
for spectra containing a low number of counts.

We have developed the tool LYNX that employs the forward fitting approach
to infer incident astrophysical spectra. However, it differs from the
conventional deterministic tools, such as XSPEC, in that the mirror and detector
characteristics are determined by incorporating Monte-Carlo simulators.

In particular LYNX links to the raytrace tool MARX (Wise et al. 1997)
to simulate the mirror response and to the PSU ACIS simulator 
(Townsley at al. 2000, in preparation)
to provide the CCD response.
Astrophysical spectra obtained with ACIS are initially fit
with the standard X-ray spectral fitting package XSPEC
to provide an initial guess for LYNX.
The modeled incident spectrum is propagated through the
Chandra mirrors and ACIS components with the MARX and the PSU ACIS simulators
respectively. A merit function that incorporates the differences between the
observed and simulated spectra spectrum is minimized
using a Downhill Simplex Method to yield the best-fit model parameters.
LYNX simulates the propagation of individual photons through the HRMA/ACIS
configuration and also takes into account the possible overlap of the
resulting charge clouds within each exposure. The 
spectra produced through LYNX therefore will simulate pile-up.
The present version of LYNX also allows fitting spectra
of any grade selection, corrects for vignetting and
accounts for the dither motion of the source across the CCD. \\

\newpage

\scriptsize
\begin{center}
\begin{tabular}{llllllllllll}
\multicolumn{12}{c}{TABLE 1} \\
\multicolumn{12}{c}{CHANDRA Observations of PKS~0637-752} \\
& & & & & & & & & & &  \\ \hline\hline
\multicolumn{1}{l} {Observation} &
\multicolumn{1}{c} {Obsid} &
\multicolumn{1}{c} {Exposure} &
\multicolumn{1}{c} {Frame} &
\multicolumn{1}{c} {SIM$_{X}$$^{a}$} &
\multicolumn{1}{c} {Src$_{Y}$$^{b}$} &
\multicolumn{1}{c} {Src$_{Z}$$^{c}$} &
\multicolumn{1}{c} {HPR$^{d}$} &
\multicolumn{1}{c} {N$_{Core}$$^{e}$} & 
\multicolumn{1}{c} {N$_{In}$$^{f}$} &
\multicolumn{1}{c} {N$_{Out}$$^{g}$} &
\multicolumn{1}{c} {N$_{Bkg}$$^{h}$} \\
Date&          &                    & Time     &           &      &      &       &      &     &                          & \\
&          &                    &          &           &      &      &       &      &     &                          & \\
&          &      s             &   s      &  mm       &arcmin&arcmin&arcsec &cnts  &cnts &cnts&cnts arcsec$^{-2}$  \\ \hline
1999-08-14T10:49:39&1051      &  1034.             & 3.24     &  0.0      & 2.0  & 1.0  &1.05   &408   &5    & 28                       & 0.028 \\
1999-08-14T11:36:00&1052      &  1041.             & 3.24     &  0.0      & 2.0  & 1.0  &0.95   &391   &2 & 28    & 0.045\\
1999-08-14T13:53:09&62558     & 19116.             & 3.24     &  1.0      & 0.0  & 0.0  &1.00   &7196  &63& 502   & 0.498\\
1999-08-14T19:15:24&62556     &  5068.             & 3.24     &  0.9      & 0.0  & 0.0  &0.95   &1799  &18& 119   & 0.111\\
1999-08-14T20:51:39&62555     &  5138.             & 3.24     &  0.65     & 0.0  & 0.0  &0.80   &1554  &18& 155   & 0.108\\
1999-08-14T22:24:18&62554     & 11367.             & 3.24     &  0.25     & 0.0  & 0.0  &0.55   &2664  &40& 281   & 0.267\\
1999-08-15T02:49:29&62553     &  5232.             & 1.541    &  0.0      & 0.0  & 0.0  &0.40   &2098  &16& 188   & 0.110 \\
1999-08-15T04:18:30&62552     &  5213.             & 1.541    & -0.5      & 0.0  & 0.0  &0.70   &2373  &20& 148   & 0.100\\
1999-08-15T05:55:56&62551     &  5343.             & 1.541    & -0.25     & 0.0  & 0.0  &0.40   &2108  &18& 148   & 0.108\\
1999-08-15T07:36:04&62550     &  5469.             & 1.541    &  0.2      & 0.0  & 0.0  &0.50   &2266  &33& 128   & 0.099\\
1999-08-15T09:12:04&62549     &  6426.             & 1.541    &  0.5      & 0.0  & 0.0  &0.55   &2777  &22& 152   & 0.133\\
1999-08-16T06:05:37&1055      &  2036.             & 3.24     &  0.0      & 0.0  & 0.0  &0.85   &659   &8 & 50    & 0.057\\
1999-08-16T06:56:36&1056      &  1757.             & 3.24     &  0.0      & 1.4  & 1.4  &1.10   &744   &5 & 42    & 0.061\\
1999-08-16T17:43:28&1058      &  1757.             & 3.24     &  0.0      &-1.4  & 1.4  &1.00   &710   &7 & 41    & 0.044\\
1999-08-16T18:29:48&1059      &  1760.             & 3.24     &  0.0      &-3.0  &-2.8  &1.00   &555   &5 & 22    & 0.250  \\
1999-08-16T19:16:07&1060      &  1760.             & 3.24     &  0.0      &-1.4  &-1.4  &1.20   &781   &6 & 51    & 0.155\\
1999-08-16T20:53:19&1062      &  1760.             & 3.24     &  0.0      & 1.4  & 1.4  &1.10   &758   &9 & 54    & 0.097\\
1999-08-16T21:35:07&1063      &  1760.             & 3.24     &  0.0      & 2.8  &-2.8  &2.00   &1019  &15& 40    & 0.115\\
1999-08-20T02:26:38&472       &  5809.             & 0.941    &  1.0      & 0.0  & 0.0  &1.05   &3417  &20& 140   & 0.095\\
1999-08-20T04:54:34&473       &  4678.             & 0.941    &  0.11     & 0.0  & 0.0  &0.4    &2206  &21& 117   & 0.085\\
1999-08-20T06:17:37&474       &  4856.             & 0.941    & -0.09     & 0.0  & 0.0  &0.4    &2255  &21& 112   & 0.088\\
1999-08-20T07:54:18&475       &  4856.             & 3.24     & -0.19     & 0.0  & 0.0  &0.45   &2386  &22& 100   & 0.088\\
1999-08-20T09:30:57&476       &  4856.             & 0.941    & -0.98     & 0.0  & 0.0  &1.10   &2967  &10& 113   & 0.167\\

\hline \hline
\end{tabular}
\end{center}

NOTES:\\
{}$^{a}$  SIM$_{X}$ Is the distance along the optical axis of the scientific instrument module
from the best focus location. \\
{}$^{b}$ Src$_{Y}$ is the distance along the Y direction from the nominal aim-point.
Y is the  direction of grating dispersion. The ACIS-S array nominal aim-point falls on chip S3, 2.0arcmin to
the right (+Y) of the edge of the chip. \\
{}$^{c}$ Src$_{Z}$ is the distance along the Z direction from the nominal aim-point.
Z is the direction normal to the grating dispersion. \\ 
{}$^{d}$ HPR is the half power radius. \\
{}$^{e}$ $N_{Core}$ are the detected events from the core component 
extracted from circles centered on the core with radii of 5$''$ \\
{}$^{f}$ $N_{In}$  are the detected events from the inner-jet component 
extracted from rectangular regions having the
left lower corners set at (x$_{c}$ + 4$''$,y$_{c}$-1.5$''$)
and the right upper corners set at (x$_{c}$ + 6.5$''$,y$_{c}$+1.5$''$),
where x$_{c}$, y$_{c}$ are the centroid locations of the core
for each observation. \\
{}$^{g}$ $N_{Out}$ are the detected events from the outer-jet component
extracted from circles centered on the X-ray knot WK8.9 with radii of 2.5$''$ \\
{}$^{h}$ $N_{Bkg}$ are the detected background events per arcsec$^{2}$ extracted 
from annuli centered on the core with inner and outer radii of 45$''$ and 55$''$ respectively.
Only events with standard ASCA grades 0,2,3,4,6 were extracted. \\

\newpage

\scriptsize
\begin{center}
\begin{tabular}{llcccccc}
\multicolumn{8}{c} {TABLE 2} \\
\multicolumn{8}{c} {Model Parameters Determined from Spectral Fits to the} \\
\multicolumn{8}{c} { {\it Chandra} ACIS-S Spectra of the Core Component of PKS~0637-752} \\
& & & & & & & \\ \hline \hline
\multicolumn{1}{c} {Fit $^{a}$} &
\multicolumn{1}{c} {Range} &
\multicolumn{1}{c} {$\Gamma$} &
\multicolumn{1}{c} {$N_{H}(z=0)$} &
\multicolumn{1}{c} {Flux $^{b}$} &
\multicolumn{1}{c} {Flux Density at 1~keV} &
\multicolumn{1}{c} {$L_{X}$ $^{c}$} &
\multicolumn{1}{c} {$\chi^{2}_{\nu}/(dof)$} \\
  &keV    &                       &$10^{20}$$cm^{-2}$      &10$^{-12}$ erg s$^{-1}$ cm$^{-2}$ &10$^{-13}$ erg s$^{-1}$ cm$^{-2}$ keV$^{-1}$& 10$^{45}$ erg s$^{-1}$ & \\ \hline
1 &0.2-6.0&1.77$_{-0.1}^{+0.1}$   &11$_{-2.0}^{+2.0}$      &1.1(2.2)                        & 7.4                                               &5.9(6.4)               & 1.4(86)  \\
2 &0.2-6.0&1.69$_{-0.05}^{+0.05}$ &3.5$_{-0.5}^{+0.5}$     &1.3(2.2)                              & 7.8                                         &4.8(6.0)               & 1.1(235)  \\
3 &0.2-6.0&2.00$_{-0.04}^{+0.04}$ &9.0(fixed)              &1.3(1.7)                              & 8.5                                         &7.8(5.5)               & 1.7(236)  \\
4 &1.0-6.0&$1.76_{-0.08}^{+0.08}$ &9.0(fixed)              &1.1(2.2)                              & 7.5                                         &5.6(6.1)               & 1.1(128)  \\

\hline \hline
\end{tabular}
\end{center}
\noindent

NOTES-\\
$ ^{a}$ All fits incorporate a power-law plus absorption due to cold material at
solar abundances. Spectral fit 1  
was performed with the LYNX spectral fitting tool 
with events extracted from node 1 only. 
Spectral fits 2, 3, and 4 were performed with the XSPEC tool
with events extracted from nodes 0 and 1.\\

$ ^{b}$ Fluxes calculated in the ranges 0.2-2~keV and 2-10~keV (quoted in parentheses).
X-ray fluxes are not corrected for Galactic absorption.\\

$ ^{c}$ Luminosities calculated in the ranges 0.2-2~keV and 2-10~keV (quoted in parentheses).
Luminosities are in the rest frame and are corrected for Galactic absorption.\\

\newpage

\scriptsize
\begin{center}
\begin{tabular}{lll}
\multicolumn{3}{c}{TABLE 3}\\
\multicolumn{3}{c}{{\it Chandra} ACIS Instrumental Spectral Features } \\
& & \\ \hline\hline
\multicolumn{1}{c} {Energy} &
\multicolumn{1}{c} {Spectral Feature$^{a}$} &
\multicolumn{1}{c} {Origin} \\
keV               &                                             &                      \\ \hline
0.076             & Al-L Absorption Edge                     & ACIS-OBF     \\
0.105             & Si L$_{3}$ Absorption Edge               & ACIS-CCD                     \\           
0.107             & Si L$_{2}$ Absorption Edge               & ACIS-CCD                     \\
0.158             & Si L$_{1}$ Absorption Edge               & ACIS-CCD                     \\
0.285             & C K${\alpha}$ Absorption Edge            & ACIS-OBF             \\
0.402             & N K${\alpha}$ Absorption Edge            & ACIS-OBF, ACIS-CCD             \\
0.535             & O K${\alpha}$ Absorption Edge            & ACIS-OBF, ACIS-CCD  \\
1.486             & Al K$_{\alpha}$ Fluorescence Line        & ACIS-CCD  \\ 
1.559             & Al K$_{\alpha}$ Absorption Edge          & ACIS-OBF             \\
1.739             & Si K Flourescence Line                   & ACIS-CCD              \\
E$_{0}$-E$_{f}$$^{b}$   & Si K Escape Peak                         & ACIS-CCD              \\
1.841             & Si K Absorption Edge of Polysilicon      & ACIS-CCD             \\
1.8473            & Si K Absorption Edge of SiO$_{2}$        & ACIS-CCD             \\
1.8447            & Si K Absorption Edge of Si$_{3}$N$_{4}$  & ACIS-CCD             \\
2.085$^{c}$       & Ir M-V Edge                              & HRMA  \\
2.112             & Au M${\alpha}_{1,2}$ Fluorescence Line   & ACIS-CCD              \\
2.156$^{c}$       & Ir M-IV Edge                             & HRMA  \\
2.20              & Au M${\beta}$ Fluorescence Line          & ACIS-CCD              \\
2.410             & Au M${\gamma}$ Fluorescence Line         & ACIS-CCD              \\
2.549$^{c}$       & Ir M-III Edge                            & HRMA  \\
2.906$^{c}$       & Ir M-II Edge                             & HRMA  \\
3.183$^{c}$       & Ir M-I Edge                              & HRMA  \\
7.469             & Ni K$_{\alpha}$ Fluorescence Line        & ACIS-CCD              \\
9.71              & Au L${\alpha}_{1}$ Fluorescence Line     & ACIS-CCD             \\
11.44             & Au L${\beta}_{1}$ Fluorescence Line      & ACIS-CCD             \\
11.52             & Au L${\beta}_{2}$ Fluorescence Line      & ACIS-CCD             \\
15.2-E$_{bias}$$^{d}$ & 4096 ADU - bias level                    & ACIS-CCD             \\ 
15.2              & 4096 ADU      &   \\

\hline \hline
\end{tabular}
\end{center}
\noindent

NOTES-\\
{}$^{a}$Additional instrumental spectral features arise
due to X-ray absorption fine structure (XAF) 
produced in the CCD's and UV/Optical blocking filter. These features
extend for about a few hundred eV above each of 
absorption edges of Al-L, C-K, N-K, O-K, Al-K. \\
{}$^{b}$ E$_{0}$ is the energy of the incident photon and E$_{f}$ = 1.739
is the energy of the silicon flourescence photons.\\ 
{}$^{c}$ Mirror absorption energies from Graessle et al. 1992, Proc. SPIE, 1742. \\
{}$^{d}$ These energies are dependent on the gain and bias levels and will vary from 
chip to chip. \\

\newpage

\normalsize

\scriptsize
\begin{center}
\begin{tabular}{llcccccc}
\multicolumn{8}{c}{TABLE 4}\\
\multicolumn{8}{c}{Model Parameters Determined from Spectral Fits to the} \\
\multicolumn{8}{c}{{\it Chandra} ACIS-S Spectra of the Inner-Jet Component of PKS~0637-752} \\
& & & & & & &\\ \hline\hline
\multicolumn{1}{c} {Model $^{a}$} &
\multicolumn{1}{c} {$\Gamma$ or T$_{e}$} &
\multicolumn{1}{c} {T$_{e}$} &
\multicolumn{1}{c} {$N_{H}(z=0)$} &
\multicolumn{1}{c} {Flux $^{b}$} &
\multicolumn{1}{c} {Flux Density at 1~keV} &
\multicolumn{1}{c} {$L_{X}$ $^{c}$} &
\multicolumn{1}{c} {$\chi^{2}_{\nu}/(dof)$} \\
  &                         &keV&$10^{20}$cm$^{-2}$   &10$^{-15}$ erg s$^{-1}$ cm$^{-2}$ &10$^{-15}$ erg s$^{-1}$ cm$^{-2}$ keV$^{-1}$& 10$^{44}$ erg s$^{-1}$ & \\ \hline

1 &2.0$_{-0.2}^{+0.2}$      &&12.6$_{-0.1}^{+0.1}$ &5.9(8.6)&4.1         & 0.42(0.29)              & 1.5(14)  \\
2 &                         &2.7$_{-0.2}^{+0.2}$keV&12.0$_{-0.3}^{+0.3}$ &5.6(3.3)&3.8         & 0.29(0.19)              & 1.3(14)  \\
3 &2.27$_{-0.2}^{+0.2}$     &&9.0(fixed)           &7.8(6.8)&5.1         & 0.63(0.25)              & 1.9(8)  \\
4 &                         &2.29$_{-0.5}^{+1.0}$keV&9.0(fixed)           &7.3(3.0)&4.7         & 0.34(0.20)              & 2.0(8)  \\
\hline \hline
\end{tabular}
\end{center}
\noindent

NOTES-\\
$ ^{a}$ Fits 1 and 3 incorporate a power-law spectrum plus absorption due to cold material at
solar abundances fixed to the Galactic value. Spectral fits 2 and 4 
incorporate a Raymond-Smith thermal plasma model with the
abundance set at 0.3 of the cosmic value.
Fits 1 and 2 were performed in the observed energy range of 0.2 to 4~keV 
using LYNX. Fits 3 and 4 were performed in the observed energy range of 0.6 to 4~keV
using XSPEC.\\

$ ^{b}$ Fluxes calculated in the ranges 0.2-2~keV and 2-10~keV (quoted in parentheses).
X-ray fluxes are not corrected for Galactic absorption.\\

$ ^{c}$ Luminosities calculated in the ranges 0.2-2~keV and 2-10~keV (quoted in parentheses).
Luminosities are in the rest frame and are corrected for Galactic absorption.\\

\newpage

\scriptsize
\begin{center}
\begin{tabular}{llccccccc}
\multicolumn{8}{c}{TABLE 5}\\
\multicolumn{8}{c}{Model Parameters Determined from Spectral Fits to the} \\
\multicolumn{8}{c}{{\it Chandra} ACIS-S Spectra of the Outer-Jet Component of PKS~0637-752} \\
& & & & & & &\\ \hline\hline
\multicolumn{1}{c} {Fit $^{a}$} &
\multicolumn{1}{c} {$\Gamma$} &
\multicolumn{1}{c} {T} &
\multicolumn{1}{c} {$N_{H}(z=0)$} &
\multicolumn{1}{c} {Flux $^{b}$} &
\multicolumn{1}{c} {Flux Density at 1~keV} &
\multicolumn{1}{c} {$L_{X}$ $^{c}$} &
\multicolumn{1}{c} {$\chi^{2}_{\nu}/(d)$} \\
  &                      &keV                        &$10^{20}$cm$^{-2}$ &10$^{-13}$ erg s$^{-1}$ cm$^{-2}$ &10$^{-14}$ erg s$^{-1}$ cm$^{-2}$ keV$^{-1}$& 10$^{44}$ erg s$^{-1}$ & \\ \hline
1 &1.83$_{-0.1}^{+0.1}$  &                           &11.8 $_{-0.3}^{+0.3}$   &0.40(0.75)                            &2.8                                         & 2.3(2.3)     & 2.0(49)  \\
2 &                      &3.8$_{-0.2}^{+0.2}$        &11.8 $_{-0.1}^{+0.1}$   &0.43(0.41)                            &3.0                                         & 2.0(1.9)     & 1.9(49)  \\
3 &1.85$_{-0.08}^{+0.08}$&                           &9.0(fixed)              &0.50(0.82)                            &3.3                                         & 2.7(2.5)     & 1.3(85)  \\
4 &                      &5.6$_{-0.6}^{+1.1}$        &9.0(fixed)              &0.48(0.61)                            &3.2                                         & 1.9(2.4)     & 1.4(85)  \\
\hline \hline
\end{tabular}
\end{center}
\noindent

NOTES-\\

$ ^{a}$ Fits 1 and 3 incorporate a power-law plus absorption due to cold material at
solar abundances fixed to the Galactic value. Fits 2 and 4 incorporate a
Raymond-Smith thermal plasma model plus absorption due to cold material.
Metal abundances were held fixed at 0.3 for fits 2 and 4. 
Spectral fits 3 and 4 were performed with XSPEC in the observed energy range of 0.6~keV to 7~keV
with simultaneous fits to spectra extracted from nodes 0 and 1.
Fits 1 and 2 were performed with LYNX in the observed energy range 0.2~keV to 7~keV
with events extracted from node 1 only.\\

$ ^{b}$ Fluxes calculated in the ranges 0.2-2~keV and 2-10~keV (quoted in parentheses).
X-ray fluxes are not corrected for Galactic absorption.\\

$ ^{c}$ Luminosities calculated in the ranges 0.2-2~keV and 2-10~keV (quoted in parentheses).
Luminosities are in the rest frame and are corrected for Galactic absorption.\\

\newpage

\begin{center}
\begin{tabular}{lllllll}
\multicolumn{7}{c}{TABLE 6}\\
\multicolumn{7}{c}{Sources in the Near Vicinity of PKS~0637-752} \\
& & & & & &\\ \hline\hline
\multicolumn{1}{c} {Object} &
\multicolumn{1}{c} {X } &
\multicolumn{1}{c} {Y } &
\multicolumn{1}{c} {RA (J2000)} &
\multicolumn{1}{c} {DEC (J2000)} &
\multicolumn{1}{c} {Distance} &
\multicolumn{1}{c} {Count Rate}  \\
    & pixel   & pixel   &       &        & arcmin            &   10$^{-2}$ cnts s$^{-1}$           \\ \hline
CXO J063551.5-751528  &4149&4209&6h35m51.5s     &-75d15m28s   & 0.89&  0.71 $\pm$ 0.06\\
CXO J063538.3-751510  &3655&4007&6h35m38.3s     &-75d15m10s   & 1.25&  1.19 $\pm$ 0.08    \\
CXO J063630.3-751522  &3705&3987&6h36m30.3s     &-75d15m22s   & 2.98& 0.37 $\pm$ 0.04\\
CXO J063607.2-751906  &3897&3257&6h36m07.2s     &-75d19m06s   & 3.16& 0.27 $\pm$ 0.04\\
CXO J063635.5-751659  &4667&4053&6h36m35.5s     &-75d16m59s   & 3.25&  0.56 $\pm$ 0.05      \\
CXO J063551.2-751929  &4047&4173&6h35m51.2s     &-75d19m29s   & 3.26& 0.26 $\pm$ 0.04\\
CXO J063642.1-751649  &3019&4185&6h36m42.1s     &-75d16m49s   & 3.63&  1.24 $\pm$ 0.08    \\
CXO J063446.7-751517  &4051&3683&6h34m46.7s     &-75d15m17s   & 4.01& 0.30 $\pm$ 0.04\\
CXO J063431.6-751627  &3925&3729&6h34m31.6s     &-75d16m27s   & 4.85&  0.64 $\pm$ 0.06\\
CXO J063500.3-752006  &3745&4185&6h35m00.3s     &-75d20m06s   & 4.90& 0.29 $\pm$ 0.04\\
CXO J063522.0-752107  &4275&3483&6h35m22.0s     &-75d21m07s   & 5.17& 0.36 $\pm$ 0.04\\
CXO J063438.0-751921  &4441&3605&6h34m38.0 s    &-75d19m21s   & 5.40& 0.23 $\pm$ 0.03\\

\hline \hline
\end{tabular}
\end{center}

NOTES-\\
${}^{a}$ Distance from core of PKS~0637-752.\\

\newpage

\begin{center}
\begin{tabular}{llll}
\multicolumn{4}{c}{TABLE 7}\\
\multicolumn{4}{c}{Radio Spectral Indices and Flux Densities of PKS~0637-752 Components}\\
& & & \\ \hline\hline
\multicolumn{1}{c} {Component} &
\multicolumn{1}{c} {4.8~GHz Flux Density} &
\multicolumn{1}{c} {8.6~GHz Flux Density} &
\multicolumn{1}{c} {${\alpha_{E}}$} \\
                   &    Jy          &   Jy          &            \\ \hline
Core               &6.373           &6.343          & -0.01        \\
Inner West Jet     &0.321           &0.200          &  0.81 \\
Outer West Jet     &0.167           &0.095          &  0.97 \\
East Jet           &0.206           &0.110          &  1.08 \\
Total Cleaned Flux &7.03 $\pm $0.02 &6.72 $\pm $0.02&       \\
\hline \hline
\end{tabular}
\end{center}
\noindent
NOTES-\\
${}^{a}$ All radio spectral energy indices follow the convention of 
flux density $\propto$ ${\nu}^{-\alpha_{E}}$
where ${\alpha_{E}}$ = ($\Gamma$ - 1).
Note that the inner and outer West radio jet are defined as the components
of the jet before and after the bend at WK9.7. 1 $\mu$Jy is equivalent to
10$^{-29}$ erg cm$^{-2}$ s$^{-1}$ Hz$^{-1}$. \\

\newpage

\normalsize

\newpage

\normalsize

\beginrefer

\refer {{Arnaud}, K. A.}, ASP Conf. Ser. 101: Astronomical Data Analysis Software and Systems V,
1996, 5, 17 \\

\refer Biretta, J. A., Stern, C. P., \& Harris, D. E., 1991, \aj, 101, 1632 \\

\refer {{Dickey}, J. M., and {Lockman}, F. J.}, 1990, \araa, 28, 215 \\

\refer {{Elvis}, M. and {Fabbiano}, G.}, 1984, \apj, 280, 91 \\

\refer Harris, D. E., Nulsen, P. E. J., Ponman, T. J., Bautz, M.,
Cameron, R. A., David, L. P., Donnelly, R. H., Forman, W. R.,
Grego, L., Hardcastle, M. J., Henry, J. P., Jones, C., Leahy, J. P.,
Markevitch, M., Martel, A. R., McNamara, B. R., Mazzotta, P.,
Tucker, W., Virani, S. N., \& Vrtilek, J., 2000, \apjl, 530, L81 \\

\refer {{Harris}, D. E., {Carilli}, C. L., and {Perley}, R. A.}, 1994, \nat, 367, 713\\

\refer {Harris}, D. E., and {Stern}, C. P., 1987, \apj, 313, 136 \\

\refer Lawson, A. J., \& Turner, M. J. L., 1997, MNRAS, 288, 920 \\

\refer Lovell, J. E. J., Tingay, S. J., Piner, B. G., Jauncey, D. L.,
Preston, R. A., Murphy, D. W., McCulloch, P. M., Costa, M. E.,
Nicolson, G., Hirabayashi, H., Reynolds, J. E., Tzioumis, A. K.,
Jones, D. L., Lister, M. L., Meier, D. L., Birkinshaw, M.,
Chartas, G., Feigelson, E. D., Garmire, G. P., Ghosh, K. K.,
Marshall, H. L., Mathur, S., Sambruna, R. M., Schwartz, D. A.,
Tucker, W. H., Wilkes, B., \& Worrall, D. M., 2000, Proc. VSOP Symbosium. \\

\refer {{Meisenheimer}, K., {Roser}, H. -J., {Hiltner}, P. R., 
{Yates}, M. G., {Longair}, M. S., {Chini}, R. and {Perley}, R. A.}, 1989, 
\aap, 219, 63 \\

\refer Schwartz, D. A., Birkinshaw, M., Chartas, G., Feigelson, E. D.,
Ghosh, K. K., Harris, D. E., Hooper, E. J., Jauncey, D. L., Lanzetta, K. M.,
Lowell, J., Marshall, H. L., Mathur, S., Piner, G., Preston, R. A.,
Tingay, S. J., Tucker, W. H., Virani, S., Wilkes, B., and Worrall, D., 2000
submitted to \apj. \\

\refer Siebert, J., Kawai, N., $\&$ Brinkmann, W., 1999,
A \& A, 350, 25\\

\refer {{Tingay}, S. J., {Murphy}, D. W., {Lovell}, J. E. J., 
        {Costa}, M. E., {McCulloch}, P., {Edwards}, P. G.,
        {Jauncey}, D. L., {Reynolds}, J. E., {Tzioumis}, A. K.,
        {King}, E. A., {Jones}, D. L., {Preston}, R. A., {Meier}, D. L.,
        {van Ommen}, T. D., {Nicolson}, G. D. and {Quick}, J. F. H.}, 1998, \apj, 497, 594 \\

\refer {{van Speybroeck}, L. P., {Jerius}, D., {Edgar}, R. J., 
{Gaetz}, T. J., {Zhao}, P.,  and {Reid}, P. B.}, 1997, \procspie, 3113, 89 \\

\refer Wise, M. W., Davis, J. E., Huenemoerder, Houck, J. C., Dewey, D.
Flanagan, K. A., and Baluta, C. 1997,
{\it The MARX 2.0 User Guide, CXC Internal Document} 
available at http://space.mit.edu/ASC/MARX/\\

\refer {{Weisskopf}, M. C. and {O'Dell}, S. L.}, \procspie, 1997, 3113, 2 \\

\refer Yaqoob, T, George, I. M., Turner, T.J., Nandra, K., Ptak, A.,
\& Serlemitsos, P. J., 1998, ApJ, 505, L90\\

\endrefer

\clearpage

\begin{figure*}[t]
\plotfiddle{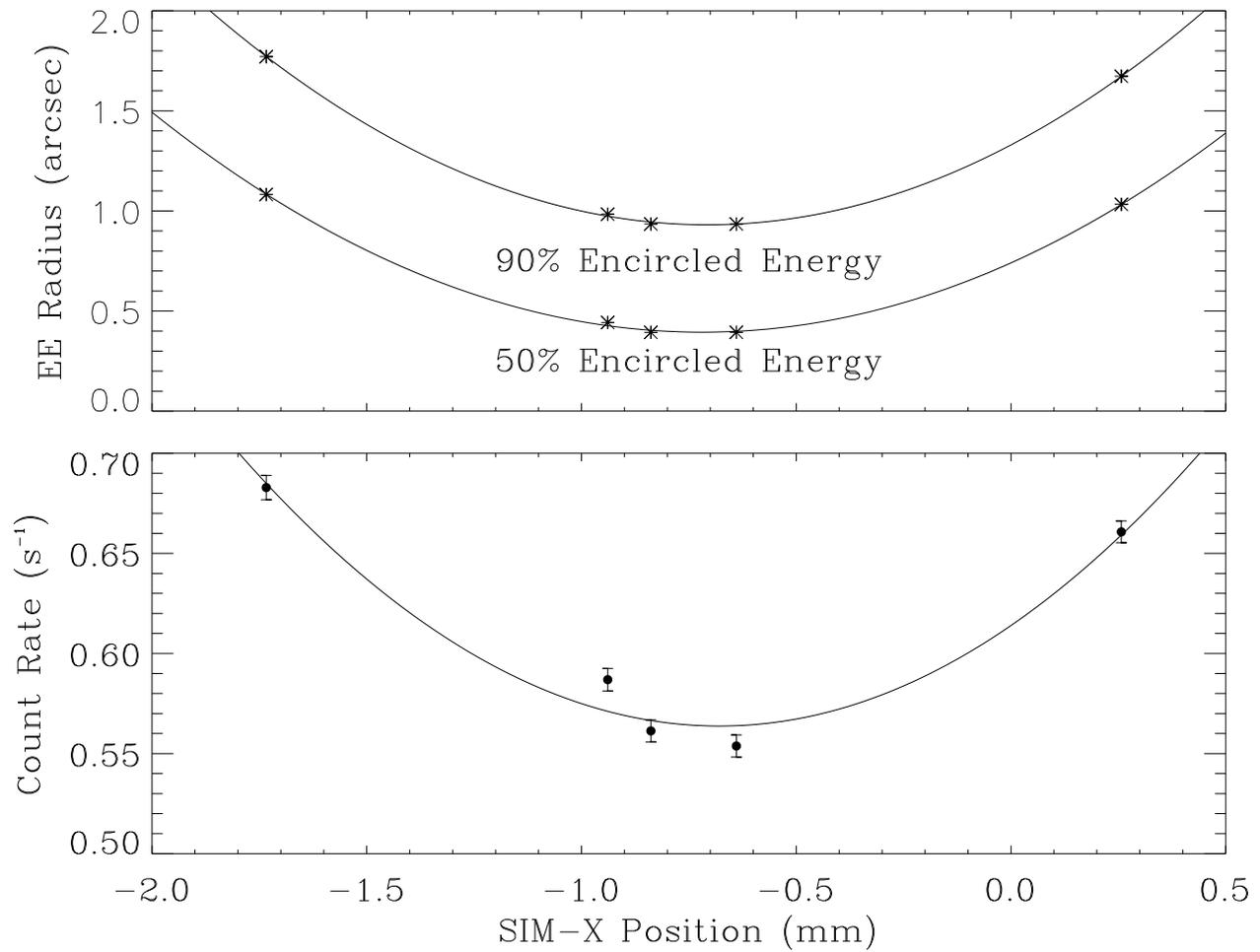}{2.in}{90}{80.}{80.}{320}{-20}
\protect\caption
{\small 50$\%$ and 90$\%$ encircled energy and observed count rate as a function of SIM X position.
The increased PSF half power radius (upper panel) and count rate
(lower panel) with SIM position away from best focus increases
because pile-up decreases.
 \label{fig:fig1}}
\end{figure*}

\clearpage

\begin{figure*}[t]
%\plotfiddle{chartas_fig2.ps}{8.5in}{0}{110.}{110.}{-310}{-170}
\plotfiddle{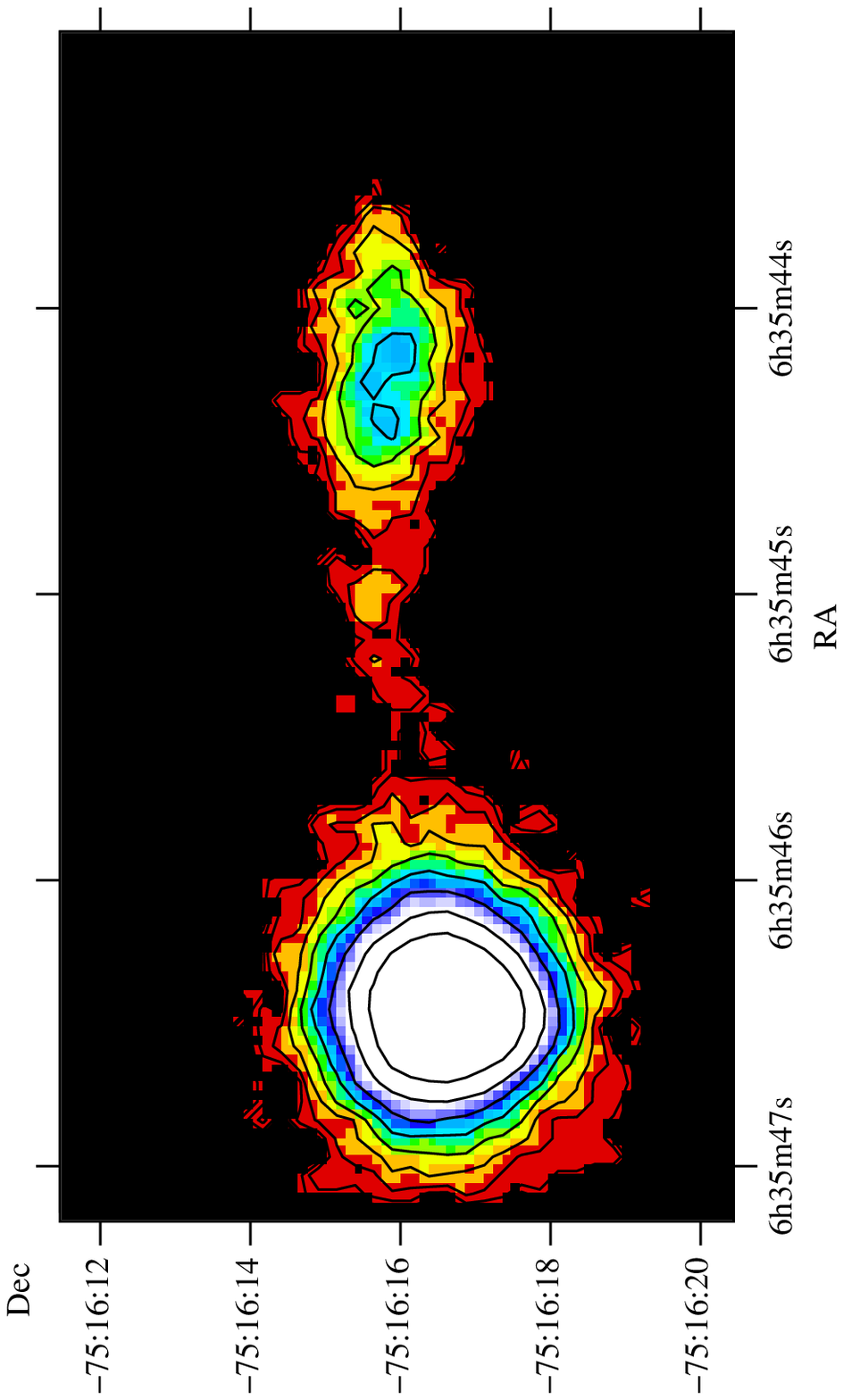}{1.5in}{-90}{60.}{60.}{-277}{328}
\plotfiddle{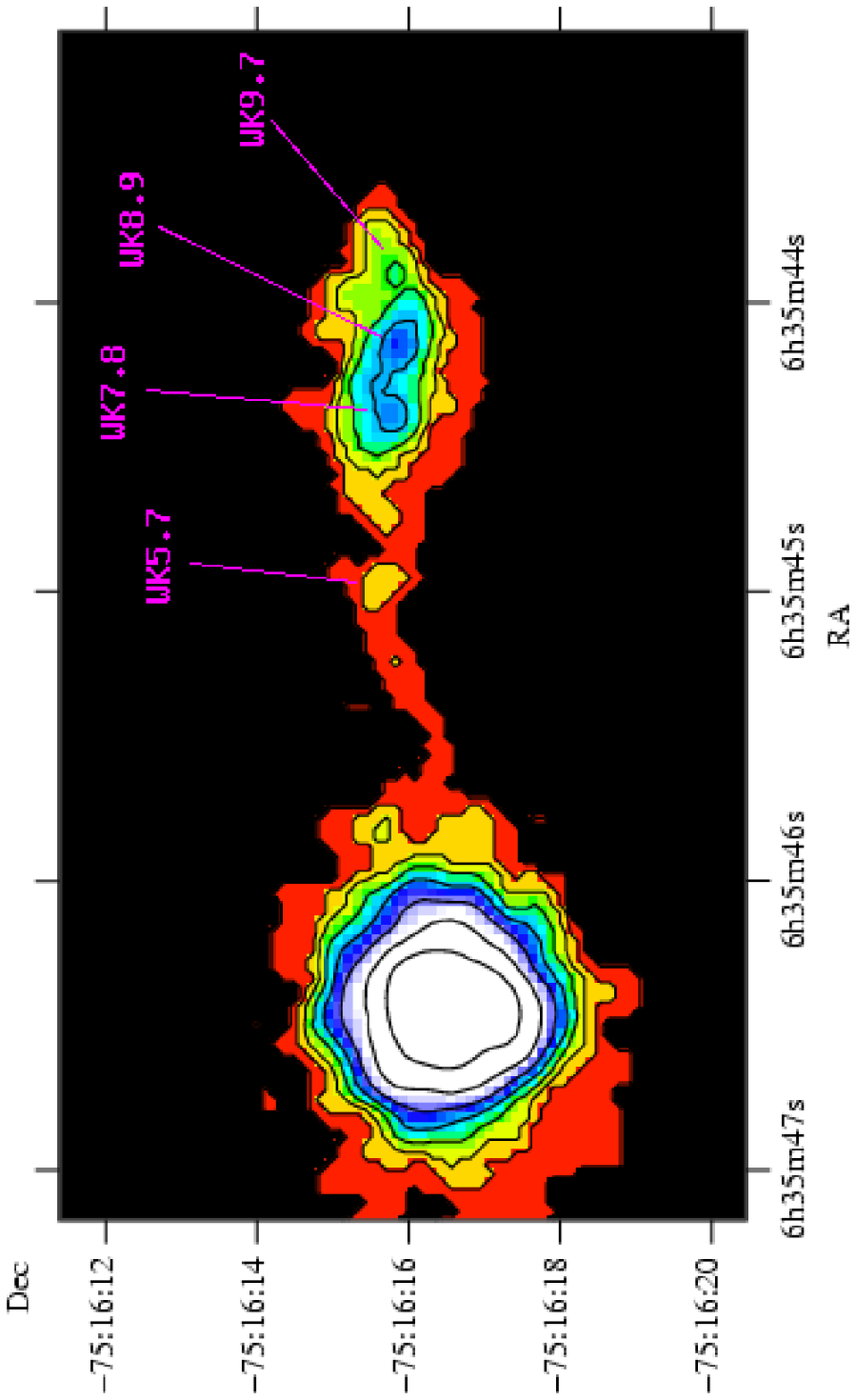}{2.45in}{-90}{60.}{60.}{-242}{321}
\plotfiddle{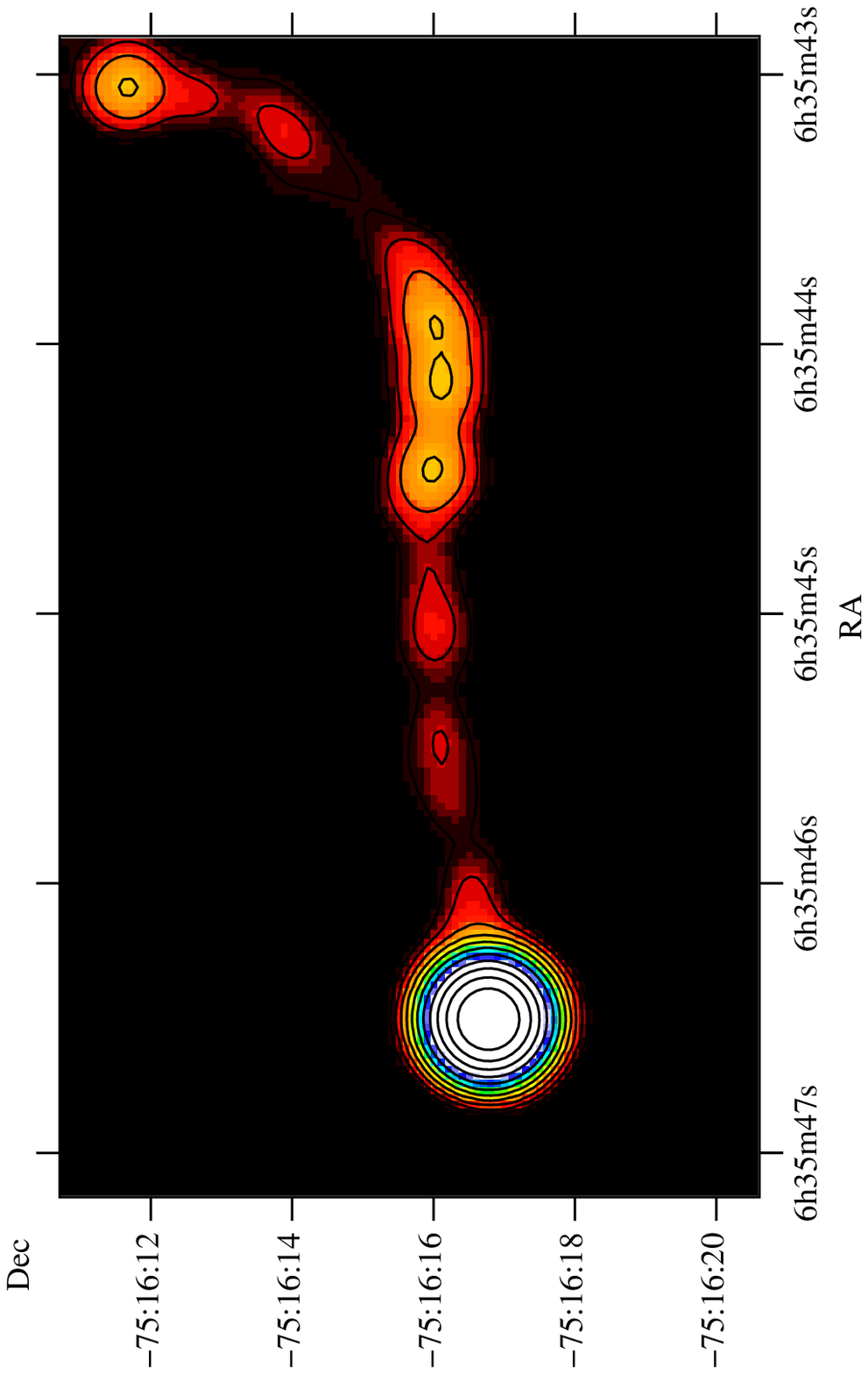}{1.5in}{-90}{64.}{64.}{-285}{293}
\protect\caption
{\small X-ray image of PKS~0637-752 created by stacking all observations 
with half power radii $<$ 1.2$''$ (Top panel).
Total maximum-likelihood deconvolved X-ray image of PKS~0637-752 produced by stacking all
deconvolved images of observations with half power radii $<$ 1.2$''$ (Middle panel).
ATCA 8.6~GHz image of PKS~0637-752 restored with a circular beam 
of 0.84$''$ FWHM (Lower panel). 
 \label{fig:fig2}}
\end{figure*}

\clearpage

\begin{figure*}[t]
\plotfiddle{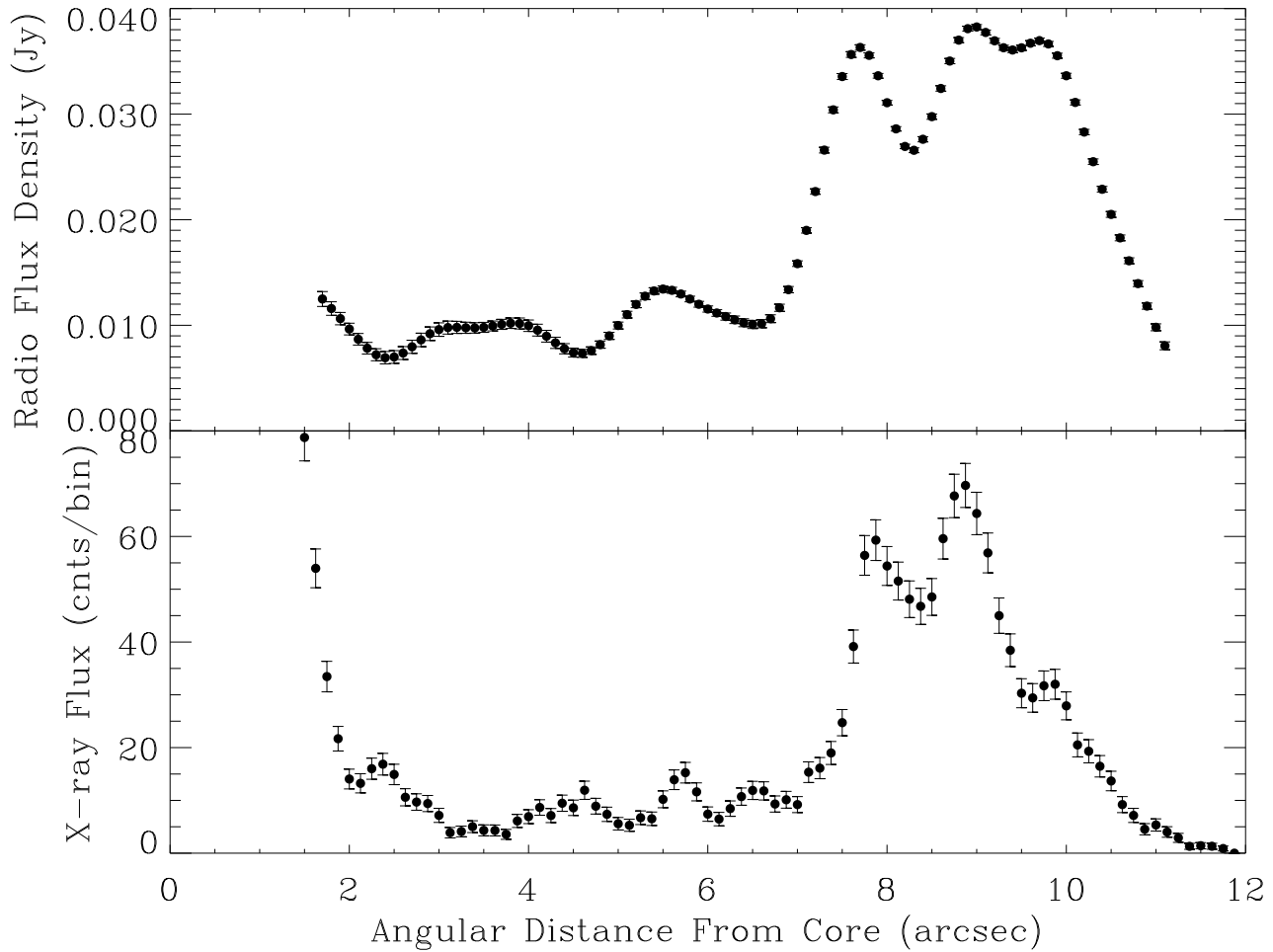}{2.in}{90}{80.}{80.}{320}{-20}
\protect\caption
{\small X-ray and radio intensity profile of jet along RA direction
integrated $\pm$ 1 arcsec perpendicular to the jet.
The X-ray profile provides counts in 0.125 arcsec increments 
and the radio profile provides the 8.6~GHz flux density per beam in 0.1 arcsec bins.
The X-ray profile was produced from the deconvolved X-ray 
image and the radio profile was produced from the 8.6~GHz 
image (Schwartz et al. 2000, in preparation; Lovell et al. 2000). 
The 8.6~GHz beam width is $\sim$ 1~arcsec FWHM,
and the effective resolution (FWHM) of the X-ray image after
deconvolution is $\sim$ 0.4~arcsec.
\label{fig:fig3}}
\end{figure*}
\clearpage

\begin{figure*}[t]
\plotfiddle{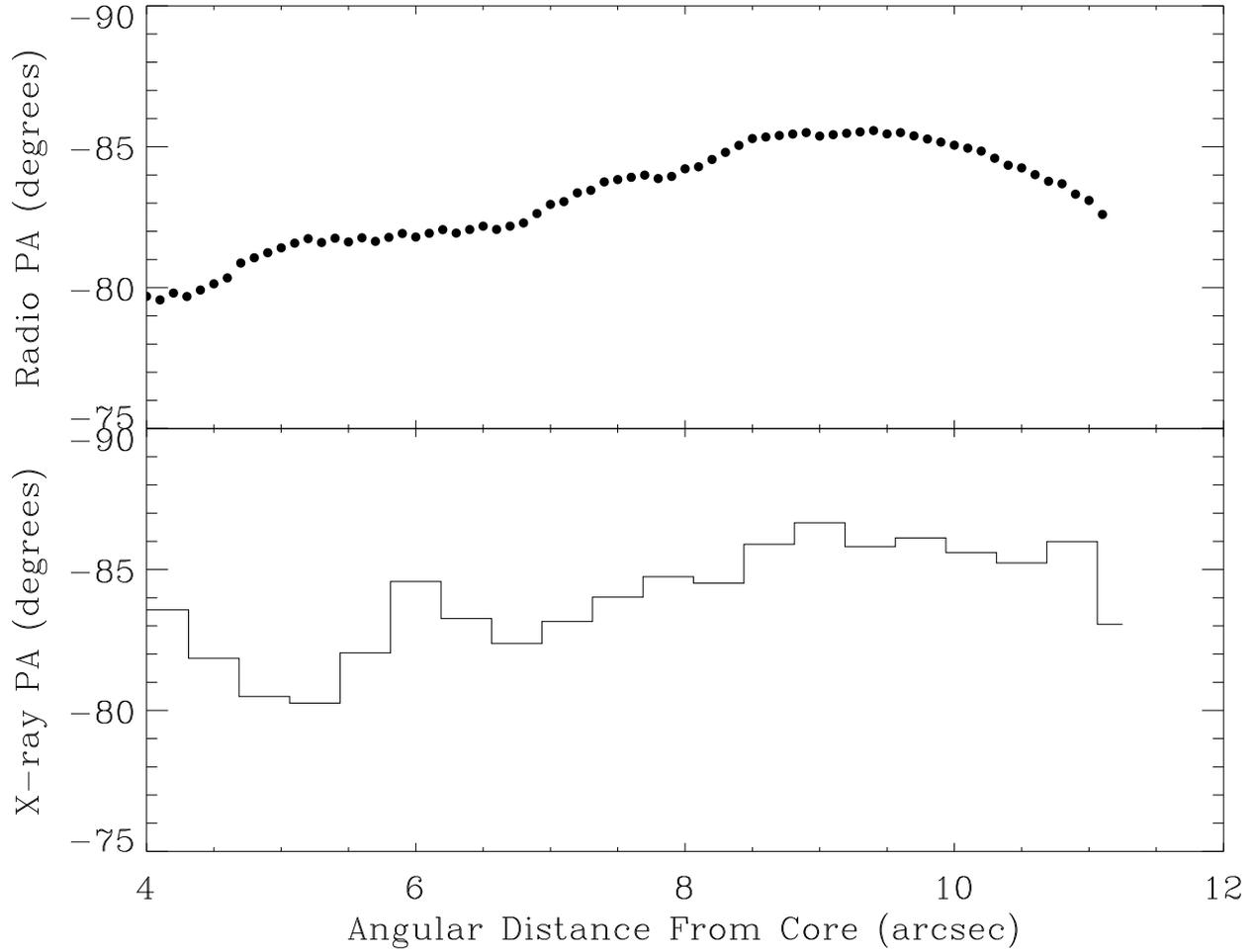}{2.in}{90}{80.}{80.}{320}{-20}
\protect\caption
{\small X-ray and radio position angle of the ridge of peak brightness along
the jet with respect to the core of PKS~0637-752. The radio position angle plot was derived from
the 8.6~GHz image (Schwartz et al. 2000, in preparation; Lovell et al. 2000).
 \label{fig:fig4}}
\end{figure*}

\clearpage

\begin{figure*}[t]
\plotfiddle{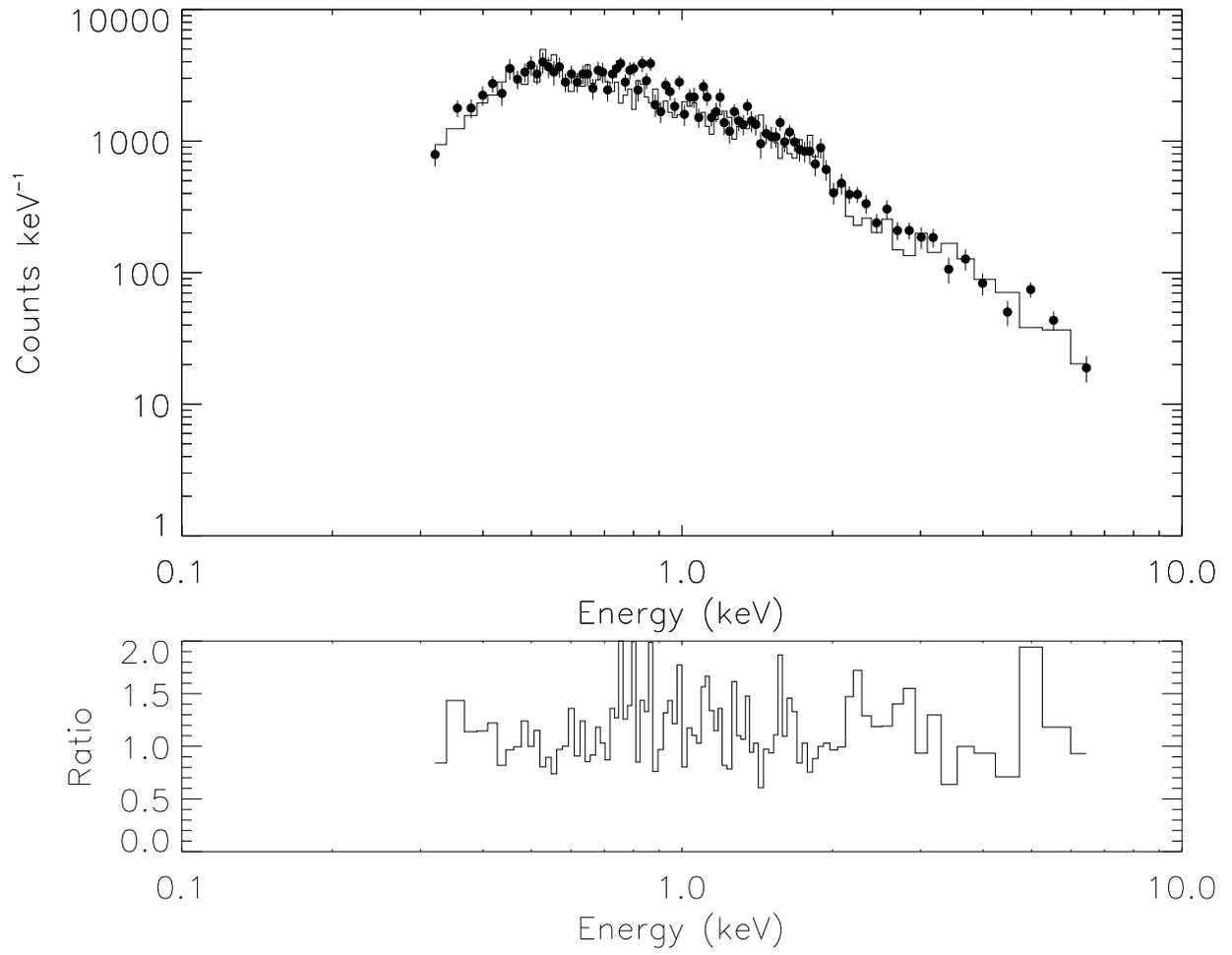}{2.in}{90}{80.}{80.}{320}{-20}
\protect\caption
{\small Spectrum from obsid 476 of core component of PKS~0637-752 with best fit model (fit 1 from Table 2).
 \label{fig:fig5}}
\end{figure*}

\clearpage
\begin{figure*}[t]
\plotfiddle{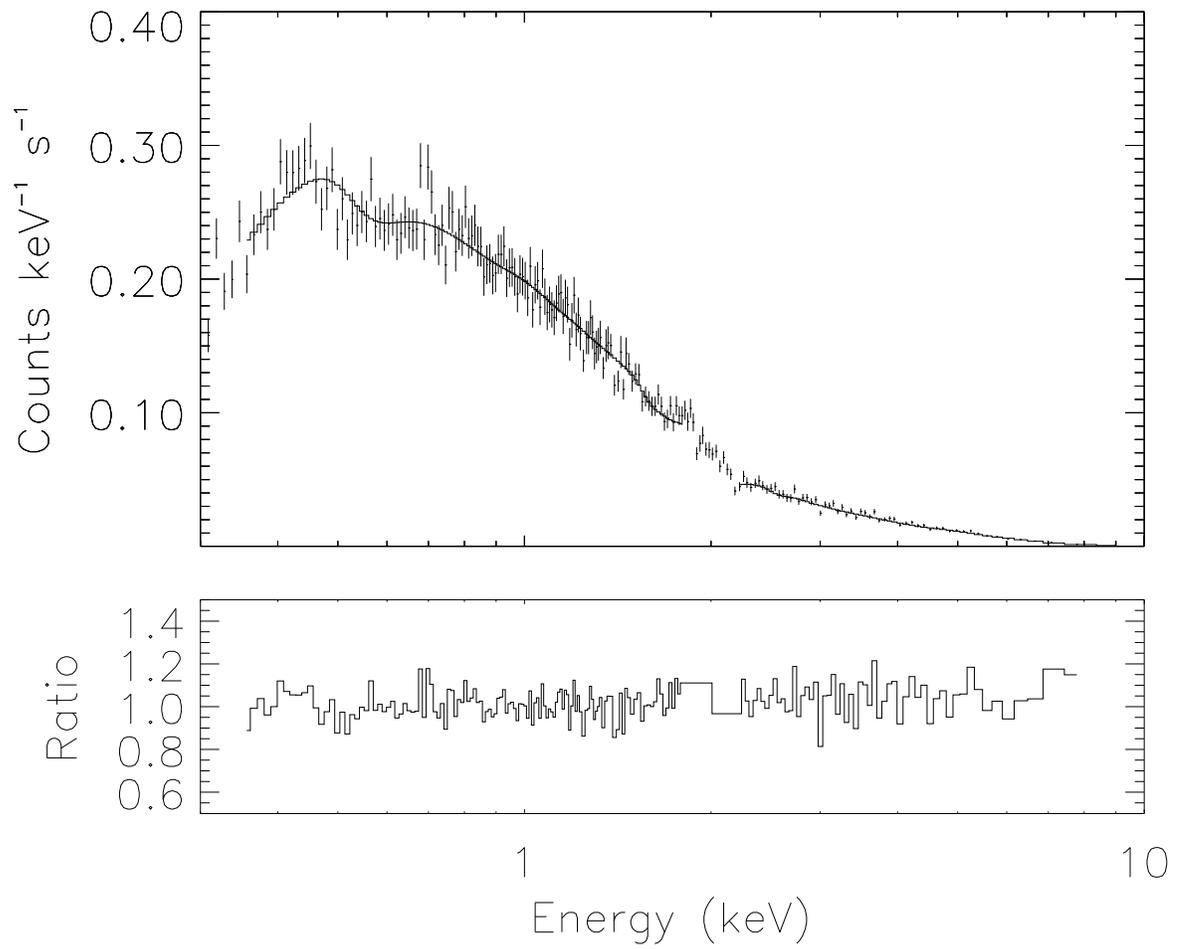}{2.in}{90}{80.}{80.}{320}{-20}
\protect\caption
{\small The stacked spectrum of the core component of PKS~0637-752. No prominent emission line
is detected near 1~keV as claimed in a recent ASCA observation of PKS~0637-752.
 \label{fig:fig6}}
\end{figure*}

\clearpage
\begin{figure*}[t]
\plotfiddle{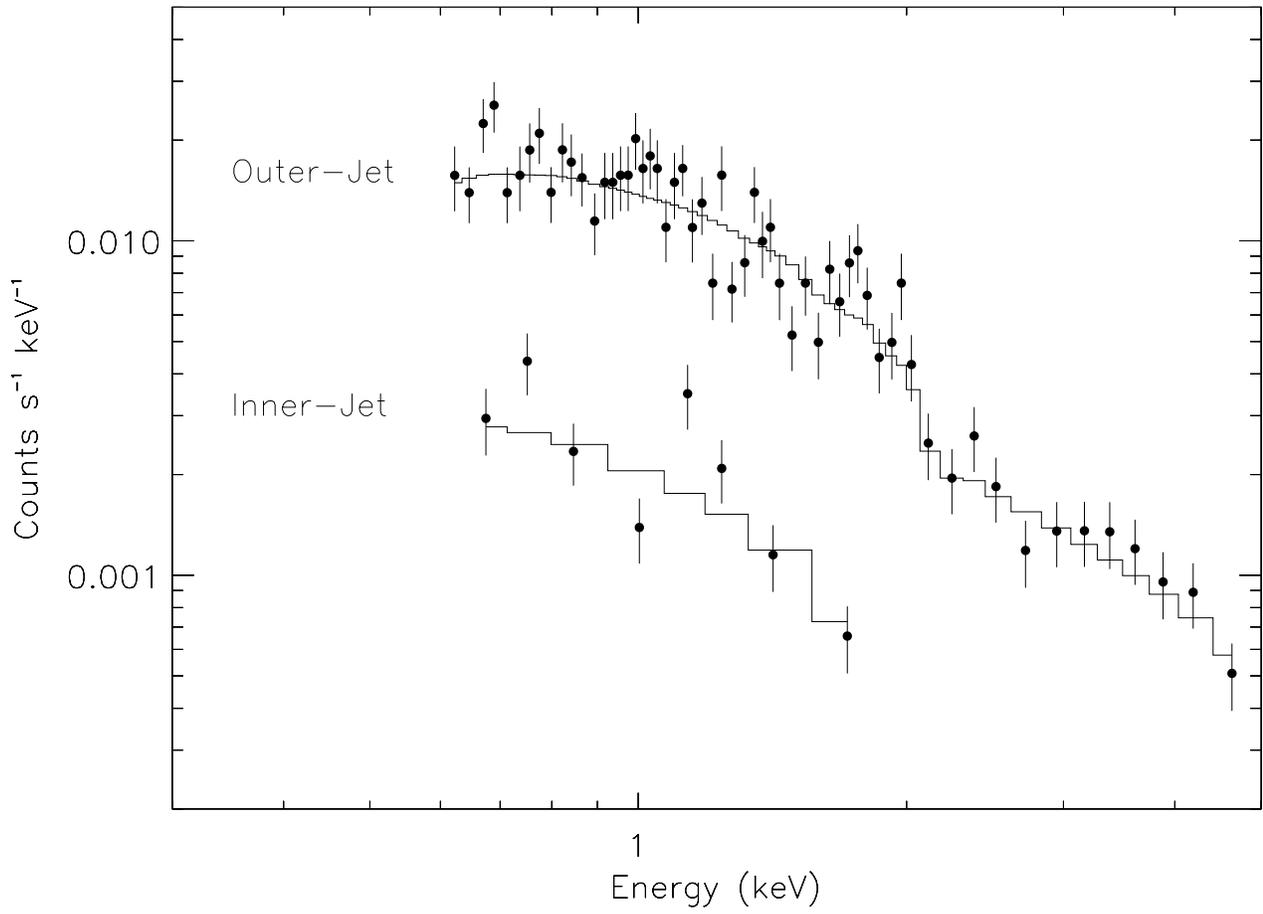}{2.in}{90}{80.}{80.}{320}{-20}
\protect\caption
{\small Spectra and best fit models of the inner (fit 3 from Table 4) and outer jet
(fit 3 from Table 5) components of PKS~0637-752.
 \label{fig:fig7}}
\end{figure*}

\clearpage
\begin{figure*}[t]
\plotfiddle{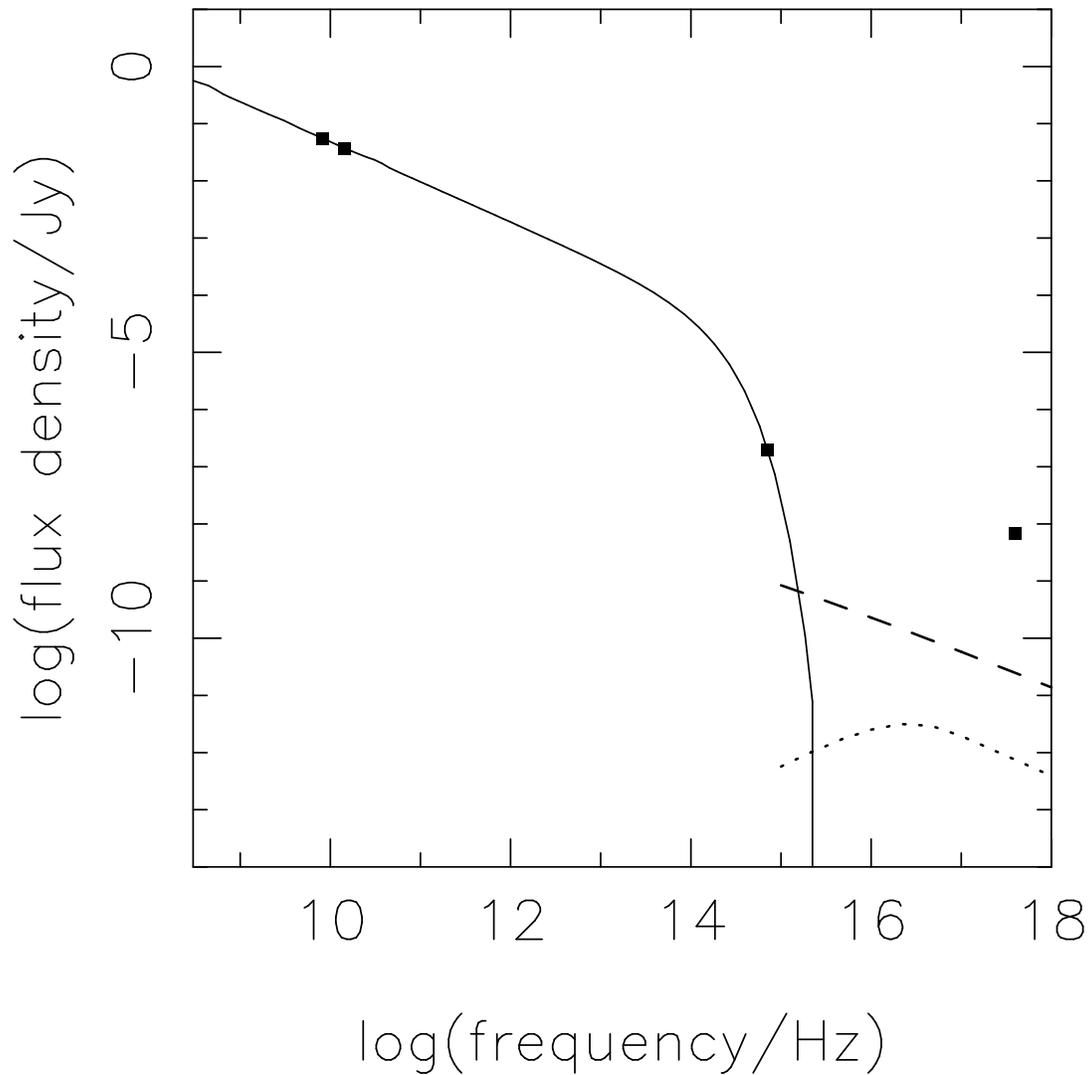}{3.in}{0}{90.}{90.}{-250}{0}
\protect\caption
{\small 
Spectral energy distribution
for knot WK7.8 in the PKS~0637-752 jet. The solid line is
the synchrotron component, the dashed line is the SSC component,
and the dotted line is the Compton-scattered CMB
component, as discussed in Section 3 and Schwartz et al.~(2000), in preparation.
The model components are based on equipartion assumptions.
 \label{fig:fig8}}
\end{figure*}

\end{document}